\documentclass[traditabstract]{aa} 
 \usepackage{longtable,lscape}

\usepackage{graphicx}
\usepackage{txfonts}
\def\gsim{~\rlap{$>$}{\lower 1.0ex\hbox{$\sim$}}}

\begin{document}

   \title{Nature vs. nurture in the low-density environment: 
 structure and evolution of early-type dwarf galaxies in poor groups\fnmsep\thanks{Based on observations obtained at the European Southern Observatory, La Silla, Chile.}
}
\titlerunning{The roles of nature and nurture in the evolution of early-type dwarf galaxies}

   \author{ F. Annibali$^1$, R. Gr\"utzbauch$^2$, R. Rampazzo$^1$, A. Bressan$^{1,4,5}$,  W. W. Zeilinger$^3$
          }

\authorrunning{F. Annibali et al.}

   \institute{INAF- Osservatorio Astronomico di Padova, 
              Vicolo dell'Osservatorio 5, I - 35122 Padova, ITALY\\
              \email{francesca.annibali@oapd.inaf.it alessandro.bressan@oapd.inaf.it roberto.rampazzo@oapd.inaf.it}
         \and
School of Physics \& Astronomy, University of Nottingham, University Park, NG7 2RD, UK \\
\email{ruth.grutzbauch@nottingham.ac.uk}
         \and
Institut f\" ur Astronomie der Universit\" at  Wien, T\" urkenschanzstra$\ss$e 17, A-1180 Wien, Austria\\
\email{werner.zeilinger@univie.ac.at}
\and
SISSA/ISAS, via Beirut 2-4, 34151 Trieste, Italy
\and 
INAOE, Luis Enrique Erro 1, 72840 Tonantzintla, Puebla, Mexico 
             } 

   \date{Received ; Accepted}

 
  \abstract
{ We present the stellar population properties of 13 dwarf galaxies residing in poor groups (low-density environment, LDE) observed with VIMOS@VLT.
Ages, metallicities, and [$\alpha$/Fe] ratios were derived within an r $<$ r$_e$/2 aperture from the Lick indices  H$\beta$, Mgb, Fe5270 and Fe5335 through comparison with our simple stellar population (SSP) models accounting for variable [$\alpha$/Fe] ratios.
For a fiducial subsample of 10 early-type dwarfs we derive median values and scatters around the medians of $5.7 \pm4.4$ Gyr, $-0.26\pm0.28$, and  $-0.04\pm0.33$ for age, $\log Z/Z_{\odot}$, and $[\alpha/Fe]$, respectively. 
 For a selection of bright early-type galaxies (ETGs) from the Annibali et al.~2007 sample residing in comparable environment
we derive median values of $9.8 \pm4.1$ Gyr, $0.06\pm0.16$, and  $0.18\pm0.13$
for the same stellar population parameters.
It follows that dwarfs  are on average younger, less metal rich, and less enhanced in the $\alpha$-elements than giants,
in agreement with the extrapolation to the low mass regime of the scaling relations derived for giant ETGs. 
From the total (dwarf$+$giant) sample we derive that 
age $\propto \sigma^{0.39\pm0.22}$, $Z\propto \sigma^{0.80\pm0.16}$, and $\alpha$/Fe $\propto \sigma^{0.42\pm0.22}$.
We also find correlations with morphology, in the sense that the metallicity and the [$\alpha$/Fe] ratio 
increase with the Sersic index $n$ or with the bulge-to-total light fraction B/T. 
The presence of a strong morphology-[$\alpha$/Fe] relation appears to be in contradiction to the possible evolution along 
the Hubble sequence from low B/T (low $n$) to high B/T (high $n$) galaxies.
We also investigate the role played by environment comparing the properties of our LDE dwarfs with those of Coma red passive dwarfs from the literature. We find possible evidence that LDE dwarfs experienced more prolonged star formations than Coma dwarfs, however 
larger data samples are needed to draw more firm conclusions.}


   \maketitle
%

\section{Introduction}

According to the widely accepted Lambda Cold Dark Matter ($\Lambda$CDM) scenario 
for structure formation, dwarf-size dark matter haloes are the first to form, 
while larger mass haloes build up later through a continuous ``hierarchical'' merging of smaller haloes 
(e.g., White \& Rees~\cite{Whi78}; Kauffmann~\cite{Kau96}). 
In this scenario, dwarf galaxies should be the first systems to form. 
However, there is strong observational evidence, both in the nearby  
(Bressan et al.~\cite{bress96}; Thomas et al.~\cite{Tho05}; Clemens et al.~\cite{Cle06}; Clemens et al.~\cite{Cle09}) and more distant Universe
(e.g., Cimatti et al.~\cite{cim06}; Fontana et al.~\cite{font06}), 
that massive early-type galaxies (ETGs) were the first to complete the star formation process,  
and that star formation terminated later in lower mass systems.
This ``anti-hierarchical'' behaviour of the stellar populations as opposed to the hierarchical behaviour of the 
dark matter is commonly referred to as  ``downsizing''. 
It has been shown that downsizing in star formation can be reconciled with the $\Lambda$CDM paradigm 
provided that a suitable mechanism (e.g., AGN feedback) is found to quench star formation at earlier times in more 
massive ETGs (e.g., Granato et al.~\cite{Gra04}; De Lucia et al.~\cite{DeL06}). 
Supernova heating is also increasingly effective with decreasing binding energy (decreasing galaxy halo mass) 
in slowing down the star formation and making the chemical enrichment less efficient 
(Granato et al.~\cite{Gra04}). This can explain the observed correlations between the average age, metallicity, and [$\alpha$/Fe] ratio (derived through integrated colours, indices, and spectral energy distribution (SED) fitting) 
and the galaxy velocity dispersion in ETGs (e.g., Nelan et al.~\cite{nel05}; Thomas et al.~\cite{Tho05}; Denicol{\'o} et al.~\cite{denic05}; Gallazzi et al.~\cite{gall05}; Clemens et al~\cite{Cle06}; Annibali et al.~\cite{Ann07}; Reda et al.~\cite{reda07}).

Another issue is the role played by environment on galaxy evolution. 
The influence of environment is testified by the presence of the well-known 
morphology-density relation, according to which early-type and morphologically 
undisturbed galaxies are preferentially located in high density environments 
(e.g., Oemler~\cite{Oem74}; Dressler et al.~\cite{Dre97}; Bamford et al~\cite{bam09}), 
and of the star-formation density relation (e.g., Lewis et al.~\cite{Lew02}; G\'omez et al.~\cite{Gom03}; 
Haines et al.~\cite{Hai07}).
This is in agreement with the  $\Lambda$CDM paradigm, where present-day 
clusters of galaxies form from the highest peaks in the primordial density fields, 
leading to an earlier onset of the collapse of the dark matter haloes and to 
more rapid mergers (e.g., Kauffmann~\cite{Kau95}).
On the other hand, studies of the stellar populations in ETGs (via absorption-line indices or 
the fundamental plane) provide discrepant results.
Some of them 
reveal a significant dependence of the stellar population properties on the local density (e.g., Longhetti et al.~\cite{long2000}; Poggianti et al.~\cite{pog01a}; Kuntschner et al.~\cite{kunt02}; Proctor et al.~\cite{proct04}; Thomas et al.~\cite{Tho05}; Clemens et al.~\cite{Cle06}; S{\'a}nchez-Bl{\'a}zquez et al.~\cite{sb06}; Annibali et al.~\cite{Ann07}; Gobat et al.~\cite{gob08}; La Barbera et al.~\cite{laba10}; Rettura et al.~\cite{rett10}), with ETGs in the field being 1-3 Gyr younger than in the cluster.
Others, claim no significant difference between the field and the cluster 
(van Dokkum \& van der Marel~\cite{vando07}; di Serego Alighieri et al.~\cite{disere06}; 
Pannella et al.~\cite{pan09}; Thomas et al.~\cite{Tho10}).

The effect of environment on galaxy evolution is expected to be stronger for dwarf galaxies, because of their lower mass. 
Low-mass galaxies are more susceptible to environmental effects due to their shallow potential well that facilitates the removal of gas from the galaxies. Hydrodynamical simulations show that, while for massive galaxies ram-pressure stripping is only effective in the cores of 
rich clusters, dwarf galaxies can be completely stripped of their gas even in poor groups (e.g., 
Bekki et al.~\cite{Bek01}; Bekki et al.~\cite{Bek02}; Marcolini et al.~\cite{Mar03}).
Suffocation, ram-pressure stripping, and galaxy harassment can deprive dwarf galaxies of their gas reservoir, quenching star formation and transforming them into passive dwarf ellipticals. They can suffer from tidal truncation (e.g., Rubin et al.~\cite{Rub88}) or tidal disruption (e.g., Oh et al.~\cite{Oh95}) during their orbit through a more massive dark matter halo, like a group or cluster of galaxies. Indeed, studies of the environmental influence on galaxy properties have found that the correlations between colour or star formation rate (SFR) and local galaxy density are the strongest for low mass galaxies (Kauffmann et al.~\cite{Kau04}; 
Tasca et al.~\cite{Tas09}; Gr\"utzbauch et al.~\cite{Gru10}). Similarly, the morphology-density relation is very strong for low-mass galaxies (e.g., Binggeli, Tarenghi \& Sandage~\cite{bts90}). It was also found that the bulk of dwarf galaxies in galaxy groups and clusters are passively evolving, while, as the local density decreases, the fraction of passively evolving dwarfs drops rapidly, reaching zero in the rarefied field (e.g., Haines et al.~\cite{Hai07}). 

Characterizing the stellar populations of dwarf galaxies in different environments is fundamental 
to test the predictions of galaxy formation models.
Studies based on absorption line-strength indices were mainly preformed for early-type dwarfs in the Virgo 
(Gorgas et al.~\cite{gorg97}; Geha et al.~\cite{geha03}; van Zee et al.~\cite{vanz04}; 
Michielsen et al.~\cite{mich08}; Paudel et al~\cite{pau10}; Spolaor et al.~\cite{spo10}), Fornax (Held \& Mould~\cite{held94}; 
Spolaor et al.~\cite{spo10}) and Coma (Poggianti et al.~\cite{pog01a}; Smith et al.~\cite{smith09}; Matkovi{\'c} et al.~\cite{matk09}) clusters. 
These works revealed a dependence of the stellar population properties with the distance from the cluster 
centre, testifying a strong environmental effect on the evolution of the stellar populations. 
The first (and unique up to date) comparison between cluster and field dwarf  early-type galaxies 
was performed by Michielsen et al~(\cite{mich08}), who however found no clear difference between the two environments. 
An analysis of a sample of dwarf ellipticals by Koleva et al.~(\cite{kol09}) revealed a complex star formation history similar to that observed in  dwarf spheroidals of the Local Group. Group and cluster galaxies have similar radial metallicity gradients and star formation histories.

Here we address the problem of the role played by internal galaxy properties and environment ({\it nature} versus {\it nurture}) 
on the evolution of early-type dwarf systems.
Our dwarf galaxies are members of four poor galaxy groups (low-density environment (LDE)) studied in Gr\"utzbauch et al.~(\cite{Gru07}) and 
Gr\"utzbauch et al.(\cite{Gru09}) and span a wide range of internal properties like luminosity, morphology (Sersic index and bulge fraction), and internal velocity dispersion. We investigate the relationship between the stellar population parameters and internal galaxy properties like morphology and velocity dispersion, as well as with local galaxy density, also comparing our sample with Coma dwarfs from the literature. 

The paper is structured as follows: the dwarf galaxy sample is described in Section~\ref{The sample};
the method to derive the stellar population parameters is detailed in Section~\ref{pop_param}. We investigate correlations of age, metallicity and [$\alpha$/Fe] ratio with internal galaxy properties, like velocity dispersion and morphology, in Section~\ref{internal}, and with environment in Section~\ref{environment}. Our results are summarized and discussed in Section~\ref{discussion}.

\begin{table*}
\begin{scriptsize}
\begin{center}
\caption{Dynamical properties of the four groups. Numbers are based on galaxies within a radius of 0.5 Mpc  from the E+S pair and a $\pm$ 1000 km s$^{-1}$ velocity interval (see Gr\"utzbauch et al.~(\cite{Gru09}) for a full description of the method).}
\label{group_properties}
\begin{tabular}{lcccccccccccc}
\hline\hline
Group & Nr. of & Distance & \multicolumn{2}{c}{Optical group centre} & $v_{group}$ & Velocity & Virial & Crossing & Virial & Group & $M/L$ \\
     & members & (Modulus) & $\alpha$ (2000)& $\delta$ (2000) &  & dispersion & radius ($R_{vir}$) & time & mass & luminosity & \\
     & within 0.5 Mpc& [Mpc (mag)] & [h:m:s] & [$^\circ$:$^\prime$:$^{\prime\prime}$] & [km s$^{-1}$] & [km s$^{-1}$] & [Mpc] & [$t_c$ H$_0$] & [10$^{12}$ M$_\odot$] & [10$^{11}$ L$_\odot$] & [$M_\odot/L_\odot$] \\
\hline
RR~143  &  5 & 50.8 (33.5) & 06:47:22.7 & -64:11:57 & 3404$\pm$82 & 189$\pm$66 &  0.49$\pm$0.05  & 0.09$\pm$0.07 & 6.1$\pm$4.8 & 1.10$\pm$0.005 &  55$\pm$5  \\
RR~210  & 23 & 31.5 (32.5) & 12:06:42.8 & -29:44:27 & 2113$\pm$34 & 165$\pm$24 & 0.70$\pm$0.10   & 0.17$\pm$0.21 & 6.5$\pm$4.3  & 1.71$\pm$0.001 & 38$\pm$4 \\
RR~216  & 12 & 47.9 (33.4) & 12:25:56.4 & -39:37:55 & 3206$\pm$80 & 273$\pm$58 & 0.71$\pm$0.10 & 0.10$\pm$0.09 & 18.3$\pm$9.4  & 3.71$\pm$0.009 & 57$\pm$9 \\
RR~242  & 22 & 50.1 (33.5) & 13:20:44.1 & -43:37:00 & 3354$\pm$76 & 363$\pm$56 &  0.44$\pm$0.02  & 0.06$\pm$0.02 & 20.3$\pm$6.9 & 1.61$\pm$0.004 & 125$\pm$7 \\
\hline
\end{tabular}
\end{center}
\end{scriptsize}
\end{table*}

\section{The sample}\label{The sample}

\subsection{Sample selection}

The dwarf galaxy  sample studied here is fully described in Gr\"utzbauch et al.~(\cite{Gru07}) and 
Gr\"utzbauch et al.~(\cite{Gru09}). All galaxies are members of poor galaxy groups dominated by a bright galaxy pair composed of an elliptical (E) and a spiral (S). The pairs were initially selected from the catalogue of isolated galaxy pairs of Reduzzi \& Rampazzo~(\cite{ReRa95}), who catalogue them as RR~143, RR~210, RR~216, and RR~242. An X-ray study of the 4 pairs (Gr\"utzbauch et al.~\cite{Gru07}) showed that, in contrast to their similar optical characteristics, their X-ray properties are very different (see also Trinchieri \& Rampazzo~\cite{TR01}).  RR~143 and RR~242 show extended diffuse X-ray emission, indicative of the presence of a hot Intra Group Medium (IGM), while RR~210 and RR~216 are X-ray underluminous, with no evidence for an IGM. 
Their diverse X-ray luminosities, L$_X$/L$_B$ ratios and X-ray morphologies suggest that they have different origins or that they might be in different evolutionary stages.

Each bright galaxy pair is surrounded by a small number of less luminous galaxies forming loose groups, and ranging from 4 group members in the poorest group RR~143, to 16 in RR~242 down to an absolute magnitude of $M_R\sim-15$ (Gr\"utzbauch et al.~\cite{Gru09}). Those additional group members were detected in a $34^\prime \times 34^\prime$ field of view, corresponding to about $500 \times 500$ kpc at the pairs' distance. The faint group members were observed with VIMOS, the VIsible Multi-Object Spectrograph (Le F\`evre et al.~\cite{LeFevre03}) at the Very Large Telescope (VLT). The observations and data reduction are described as well in 
Gr\"utzbauch et al.~(\cite{Gru09}).

Table~\ref{group_properties} gives the basic kinematical and dynamical properties of the four groups, calculated with galaxies within a 0.5 Mpc radius and velocity interval of $\Delta v_{rad} = \pm 1000$km s$^{-1}$ from the central E+S galaxy pair in each group. The calculation of all quantities is fully described in 
Gr\"utzbauch et al.~(\cite{Gru09}). The columns are group ID (1); number of group members (2), including the faint galaxies used in this study, plus additional spectroscopically confirmed members within r = 0.5 Mpc not studied here; group distance (3) in Mpc and distance modulus; coordinates(4 and 5); group radial velocity (6); the group velocity dispersion (7); virial radius (8); crossing time (9); virial mass (10); group luminosity (11) and mass-to-light ratio of the group (12). The group richness ranges from 5 (RR~143) to 23 (RR~210) members within 0.5 Mpc, and is not correlated with the presence of an X-ray emitting, hot IGM. The group masses are of the order of $6 \times 10^{12}$ to $2 \times 10^{13}$ M$_\odot$.

The morphological types of the group faint members considered in this study are given in 
Col.~2 of Table~\ref{dwarf_properties}. 
The majority of the galaxies  are classified as early-type (S0 or dE). Only two galaxies out of 17 are classified as spirals.
For RR242\_25575 the classification is not clear because of a warped structure, but it 
has a strong bulge and intermediate Sersic index, indicative of early-type morphology. 
We therefore included this galaxy in the early-type sample. M$_R$ is in the range $-19.8$ -- $-14.8$ mag.

\begin{table*}
\centering
\caption{Basis properties of the sample.}\label{dwarf_properties}
\renewcommand{\footnoterule}{} 
\begin{tabular}{lcccccccc}
\hline \hline
Galaxy &  Type  & M$_R$  &  $\sigma_{r_e/2}$  & $r_e$ & B/T & $n$ & $\rho_{xyz}$ & ${\rm \log (\Sigma_3)}$  \\   
            &  & & [km s$^{-1}$]  & [kpc ('')] & & & [Mpc$^{-3}$] &  [$\Sigma_3$ in Mpc$^{-2}$ ]\\   
\hline
RR143\_9192 & SB0   &   -16.23  & 46.0 $\pm$ 7.6 & 1.3 (4.7)   & 0.77 $\pm$ 0.04 & 1.05$\pm$0.03    &  0.09 & 1.64   \\ 
RR143\_24246 & S0   &   -16.53  & 47.3 $\pm$ 4.3 & 1.4 (5.0)   & 0.00 $\pm$ 0.04 & 1.57$\pm$0.04     &  0.09 & 1.63  \\ 
RR210\_11372 & dE   &   -14.90  & 60.1 $\pm$ 11.3 & 1.0 (7.0) & 0.00 $\pm$ 0.01 & 0.83$\pm$0.07   &  1.21 & 1.91  \\ 
RR210\_13493 & SB0 &   -18.23  & 60.9 $\pm$ 1.3 & 1.3 (9.4)   & 0.23 $\pm$ 0.01 & 1.98$\pm$0.09    &  1.08 & 1.94  \\ 
RR216\_3519 & dE     &   -16.17  & 58.9 $\pm$ 14.8 & 1.2 (5.2) & 0.04 $\pm$ 0.01 & 1.02$\pm$0.05    &  0.81 & 2.05  \\ 
RR216\_4052 & SB0   &   -18.86  & 51.7 $\pm$ 2.0 & 2.5 (10.8) & 0.08 $\pm$ 0.01 & 1.07$\pm$0.07   &  0.81& 1.85  \\ 
RR216\_12209 & Sc   &   -17.85  & 59.8 $\pm$ 7.3 & 2.2 (9.3)   & 0.00  $\pm$ 0.01&  0.74$\pm$0.03     &  0.81& 2.07 \\ 
RR242\_8064 & dE     &   -15.65  & 45.2 $\pm$ 16.3 & 1.3 (5.5) & 0.29 $\pm$ 0.04 &  1.21$\pm$0.04   & 0.72 & 1.98   \\ 
RR242\_13326 & S0   &   -17.35  & 58.8 $\pm$ 10.3 & 2.1 (8.7) & 0.10 $\pm$ 0.01 & 1.42$\pm$0.01    & 0.72 & 2.70 \\ 
RR242\_15689 & dE   &   -15.77  & 55.9 $\pm$ 3.9 & 0.7 (3.0)   & 0.06 $\pm$ 0.01 & 1.38$\pm$0.01      & 0.72 & 3.12 \\ 
RR242\_20075 & dE   &   -17.17  & 67.3 $\pm$ 3.2 & 1.4 (5.7)   & 0.12 $\pm$ 0.01 & 1.51$\pm$0.01      & 0.72 & 2.35  \\ 
RR242\_22327 & SB0 &   -18.44  & 84.9 $\pm$ 2.1 & 1.0 (4.0)   & 0.41 $\pm$ 0.01 & 1.34$\pm$0.08     & 0.72 & 3.05 \\   
RR242\_23187 & S     &   -16.13  &                           & 0.9 (3.9)  & 0.57 $\pm$ 0.07 & 0.75$\pm$0.03       & 0.72 & 2.98 \\   
RR242\_24352 & SB0 &   -19.79  & 107.7 $\pm$ 2.0 & 1.7 (6.9) & 0.29 $\pm$ 0.02 & 2.35$\pm$0.09   & 0.72 & 3.11 \\   
RR242\_25575 & SB0? &   -17.08  & 50.4 $\pm$ 4.5 & 1.7 (7.1) & 0.31$\pm$ 0.06  & 2.24$\pm$0.04   & 0.72 & 2.31 \\   
RR242\_28727 & dE   &   -16.45  & 55.6 $\pm$ 3.7 & 0.7 (2.9)   & 0.38 $\pm$ 0.02  & 1.16$\pm$0.01    & 0.72  & 3.32 \\   
RR242\_36267 & dE   &   -14.83  & 39.8 $\pm$ 15.7 & 1.1 (4.4) & 0.01 $\pm$  0.01 & 0.89$\pm$0.01   &  0.72  & 2.01 \\  
\hline
\end{tabular}
\end{table*}

\subsection{Spectroscopic observations}

To obtain line-strength indices for the most prominent Balmer and metal absorption lines in the blue range of the spectrum, we used the HR (high resolution) blue grism. This grism permits the coverage of the spectral region containing the H$\beta$, Mg2, and Fe ($\lambda$ 5270 \AA, $\lambda$ 5335 \AA) absorption lines with a resolution of R=2050 ($1^{\prime\prime}$ slit) and a dispersion of $0.51$ \AA\ pixel$^{-1}$. Spectrophotometric and Lick standard-stars were either observed or extracted from the VIMOS data archive with the same instrument set-up.  This configuration allows only one slit in the dispersion direction, i.e., each single spectrum covers the full length of the detector.  The wavelength interval depends on the slit position.  At the CCD centre, the wavelength interval is 4150 -- 6200 \AA.  At the upper CCD edge ($+ 4^\prime$), the interval is 4800 -- 6900 \AA\ , and 3650 -- 5650 \AA\ at the lower edge ($-4^\prime$). The total integration time per object was 3600 seconds.

Standard data reduction was performed with the ESO data reduction pipeline using the recipes especially developed for VIMOS, yielding bias-corrected, wavelength-calibrated, and sky-subtracted 2D spectra of each object. For the redshift measurements, the 1D spectra were extracted from the full spatial extent of the galaxy, adding up the signal over the whole 2D spectrum. For the measurement of line-strength indices, however, it is important that the same area relative to a galaxy's size is sampled in each object to avoid biasing the resulting ages and metallicities.
Since radial age and metallicity gradients are commonly detected in early-type galaxies and spheroids, sampling different radial extents, e.g. because of instrumental constraints, can introduce a systematic bias.
Due to the overall faintness and low signal-to-noise ratios of our sample, we chose to extract the spectra from a relatively large aperture of radius $r_e / 2$. The effective radii were measured with the {\tt GALFIT} package 
(Peng~\cite{Peng02}) by fitting a single Sersic model to the galaxy image, as described in 
 Gr\"utzbauch et al.~(\cite{Gru07}). The 2D spectra were extracted by adding up the pixel rows across the spatial direction that are within $\pm r_e/2$ from the peak of the light distribution (i.e. the galaxy centre).

\subsection{Galaxy velocity dispersions}

The stellar velocity dispersions were measured from the 1D extracted
spectra with the penalized pixel-fitting (pPXF) code of Cappellari \&
Emsellem~(\cite{cap04}) parametrizing the line-of-sight velocity distribution
(LOSVD) as a sum of orthogonal functions in a Gauss-Hermite
series. Spectral regions contaminated by emission lines were masked in
the galaxy spectra. The velocity dispersion was derived by a linear
combination of appropriately convolved template star spectra, which
best reproduces the galaxy spectrum.  The advantage of this method is
that errors due to template-mismatch are reduced by using a
combination of template stars. The template star collection comprised
1388 stars from the ELODIE library (Prugniel \& Soubiran~\cite{pru01}).  The
resulting values, together with their formal errors, are given for each
galaxy in Table~\ref{dwarf_properties}.

\subsection{Bulge fractions and Sersic indices }

We use two quantities to describe the galaxy's morphology and structure: (1) the bulge fraction, B/T, and (2) the Sersic index $n$.  B/T is the ratio between the luminosity of the bulge component and the {\it total} light of the galaxy, and was measured with {\tt GALFIT}. The method is fully described in
Gr\"utzbauch et al.~(\cite{Gru09}). A disk and a bulge component are fitted to the galaxy image simultaneously by an exponential and a Sersic model, respectively. The bulge luminosity is then divided by the sum of the luminosities of bulge and disk components, i.e. the total galaxy light. The Sersic index $n$ was measured by Gr\"utzbauch et al.~(\cite{Gru07}) and is obtained by fitting a {\it single} Sersic model to the whole galaxy. The Sersic index $n$ determines the shape of the radial light profile, and is found to correlate with the galaxy luminosity or mass 
(e.g., Trujillo et al.~\cite{Tru04}). More massive galaxies tend to be more centrally concentrated, showing higher $n$. 
 Furthermore, bulge dominated galaxies typically have $n>2$, where $n=4$ corresponds to the de Vaucouleurs profile, which was found 
to be valid for many massive ellipticals. The presence of a disk leads to smaller $n$, since the light profile of a disk decreases exponentially with radius, meaning $n=1$.
We do not use the Sersic index of the bulge obtained by the bulge-disk decomposition, since we are interested in possible variations of the stellar population parameters with the structure of the entire galaxy. The effective radius from the same single Sersic fit is used to determine the aperture size $r_e/2$, from which the spectra are extracted, and within which the galaxy's velocity dispersion is measured (see previous section).
We provide in Cols.~5, 6 and 7 of Table~\ref{dwarf_properties} the effective radius $r_e$, the bulge fraction B/T, and the Sersic index $n$, 
respectively. 
The errors on B/T were obtained from the magnitude errors of the bulge and disk components provided by {\tt GALFIT}.
These formal errors could underestimate the real errors. In fact, Gr\"utzbauch et al.~(\cite{Gru09}) showed that there is an average difference 
of $\sim$ 0.1 in the calculated  B/T if one fits the bulge with a de Vaucouleur's  or with a Sersic profile. The real errors are more likely close to this value. The errors on the Sersic index $n$ are those given by {\tt GALFIT}. Notice however that simulations performed with 
artificial galaxies provide slightly higher errors (e.g., H{\"a}ussler et al.~\cite{haus07}).

\subsection{Local galaxy density}

To describe the environment in which our dwarf galaxies reside, we use two quantities: the volume density $\rho_{xyz}$ 
and the local surface density  $\Sigma_3$.

The density $\rho_{xyz}$ was calculated  according to the definition of Tully~(\cite{Tul88}) 
as the density of galaxies brighter than $M_B = -16$ mag.
The density was determined in a cube of 1 Mpc$^3$ around the group's barycenter.
Using velocity-based distances, the contributions of the individual 
galaxies were summed up in order to derive  the local density for each group which are
0.09, 1.21, 0.81, and 0.72 galaxies~Mpc$^{-3}$ for RR43, RR~219, RR~216 and RR~242, 
respectively. The values reported in Col.~8 of Table~\ref{dwarf_properties} for the invidual galaxies 
reflect the local density at the position of the galaxy.
The computation of  $\rho_{xyz}$ includes a correction for
incompleteness based on the predictions by the normalized Schechter~(\cite{sche76})  
luminosity function. A detailed discussion about possible errors on $\rho_{xyz}$  is given in Tully~(\cite{tully88b}).
Because of the adopted 1 Mpc scale, the $\rho_{xyz}$ values are large-scale densities, describing the 
{\it global} properties of the group;  this explains why 
the values are quite similar within the same group.
These densities are useful for comparison between different large environments like the field and the cluster environment. 
However, they do not describe environmental differences within the same group.

A more {\it local} measure of the density is provided by the surface density $\Sigma_3$, which is  based on  the distance to the third nearest neighbour.
We considered all spectroscopically confirmed group members, including the brighter galaxies that are not studied here. For a list of group members found around each galaxy pair, see Table~A.1 in  Gr\"utzbauch et al.~(\cite{Gru09}).
The $3^{rd}$ nearest neighbour density, $\Sigma_3$, is computed by dividing the number of galaxies, in this case $N=3$, by the area over which they are distributed, i.e. the area of a circle of radius $D_3$, where $D_3$ is the distance to the third nearest neighbour:  $\Sigma_3 = 3/(\pi D_3^2)$. $D_3$ is measured in Mpc, which gives $\Sigma_3$ as a surface density in Mpc$^{-2}$. The use of the distance to the $3^{rd}$ nearest neighbour can lead to relatively high values of galaxies per Mpc$^2$, since it is sensitive to very local galaxy concentrations and is not comparable to measurements of the large-scale galaxy density,  such as $\rho_{xyz}$. 
These very local galaxy concentrations, however, can have a significant influence on the stellar populations of galaxies through galaxy interactions and merging.  The derived $\Sigma_3$ are given in Col.~9 of Table~\ref{dwarf_properties}. 
We have not computed individual errors on $\Sigma_3$, since the uncertainty is dominated by projection effects which we are not able to 
account for with the present data. 

It can be seen that $\Sigma_3$ can vary significantly within the same group. 
These values will be used in Section~5.1 to investigate the effect of local environment on the evolution of the dwarf galaxy stellar populations.

\begin{table}
\caption{[OIII] emissions for the sample.}\label{o3em}
\begin{tabular}{lc}
\hline \hline
Galaxy &   EW [OIII]$\lambda$5007 \\   
 &   [\AA] \\   
\hline
RR143\_9192  &  0.7  \\ 
RR143\_24246 & 6.2 \\ 
RR210\_11372 &  \\ 
RR210\_13493 &  $<$0.1 \\ 
RR216\_3519  &  0.3 \\ 
RR216\_4052  &   $<$0.1  \\ 
RR216\_12209 &  1.6\\ 
RR242\_8064  &  \\ 
RR242\_13326 &  1.6\\ 
RR242\_15689 &  0.2\\ 
RR242\_20075 &  0.1\\ 
RR242\_22327 &  0.1\\ 
RR242\_23187 & 27.5 \\
RR242\_24352 &  0.1\\ 
RR242\_25575 & 0.6  \\ 
RR242\_28727 &  $<$0.1 \\ 
RR242\_36267 &  0.2 \\ 
\hline
\end{tabular}
\end{table}

\section{Derivation of the stellar population parameters}\label{pop_param}

\subsection{Lick Indices}

Lick indices (Worthey et al.~\cite{worth94}) were computed following the procedure described in Rampazzo et al.~(\cite{Ram05}).
The VIMOS HR spectra extracted within the central  ${\rm r<r_e/2}$ 
aperture were first degraded to match the 
wavelength-dependent resolution of the Lick-IDS  system (FWHM$\sim$ 8.4 \AA \ at 5400 \AA, 
Worthey \& Ottaviani~\cite{Wor97}) by convolution with a Gaussian kernel of variable width.
Then the indices were computed on the degraded galaxy spectra using the refined 
passband definitions of Trager et al.~(\cite{Tra98}). Only in few cases the covered spectral range allows 
sampling of H$\gamma$ and H$\delta$.  For sake of homogeneity within the sample, we did not consider these
features, and used only H$\beta$ as a Balmer line.
The ``raw'' indices were calibrated into the  Lick-IDS  system performing, in the order, the following 
steps: correction for velocity dispersion, correction of H$\beta$ for possible emission, 
calibration through Lick-IDS standard stars.

In the first step, we convolved the spectra of two K0 III stars (HD073665 and HD073710), belonging to the Lick star dataset observed 
together with the galaxies, with Gaussian kernels of variable widths, to simulate the effect of galaxy velocity dispersion on the indices. 
Lick indices were computed on the smoothed stellar spectra to derive the fractional 
index variation R$_{\sigma}=$ (EW$_{\sigma}$ - EW$_0$)/EW$_0$ as a function of $\sigma$, 
where EW$_{\sigma}$ is the index strength at a velocity dispersion of $\sigma$, and EW$_0$ is the 
index strength at zero velocity dispersion.
It is easy to demonstrate that the zero-velocity galaxy index can be computed as ${\rm EW_{0 \sigma}=EW_{old}/ (1 + R_{\sigma})}$, where R$_{\sigma}=$ is derived for each galaxy of given $\sigma$.

\begin{table}
\caption{Lick indices.}\label{indices}
\tiny{
\begin{tabular}{lcccc}
\hline \hline
Galaxy &  H$\beta$ & Mgb  &  Fe5270  &  Fe5335 \\   
& [\AA] & [\AA] & [\AA] & [\AA] \\
\hline
RR143\_09192    &    2.62     $\pm$   0.26   &   0.45   $\pm$   0.19   &      1.37  $\pm$   0.31   &    1.40  $\pm$   0.11 \\   
RR210\_13493    &   1.92     $\pm$    0.15   &   3.32   $\pm$  0.18    &      3.07  $\pm$  0.31    &    2.34  $\pm$   0.19   \\
RR216\_03519    &   1.68     $\pm$   0.35    &   2.52   $\pm$   0.37   &      2.65 $\pm$   0.48    &     3.31 $\pm$   0.43  \\
RR216\_04052   &   2.03     $\pm$   0.14    &   2.93   $\pm$   0.17   &       2.80 $\pm$  0.31    &     2.03 $\pm$   0.18   \\
RR216\_12209    &   2.32     $\pm$  0.62     &   1.29   $\pm$  0.19    &       2.03  $\pm$  0.32   &    2.00  $\pm$  0.20    \\
RR242\_08064    &   2.45     $\pm$   0.15    &  1.79    $\pm$ 0.18     &       3.01  $\pm$ 0.31    &     0.76  $\pm$ 0.21 \\
RR242\_13326 &     4.47      $\pm$  0.61    &  3.37     $\pm$ 0.13     &       2.29 $\pm$ 0.29 &   \\ 
RR242\_15689    &  1.87     $\pm$   0.17     &  2.24    $\pm$ 0.18     &       2.68 $\pm$  0.31  &     2.52  $\pm$ 0.18 \\
RR242\_20075   &  1.96      $\pm$   0.18     &  2.41    $\pm$  0.21    &       2.63  $\pm$ 0.33  &     1.58  $\pm$  0.22  \\
RR242\_22327  &  1.49       $\pm$   0.18     & 4.06     $\pm$ 0.20     &       2.78 $\pm$ 0.33   &      2.16 $\pm$ 0.22 \\
RR242\_24352    &  1.98     $\pm$  0.13      & 4.07     $\pm$  0.16    &       3.34 $\pm$  0.30    &      2.86  $\pm$  0.17  \\
RR242\_25575   &  4.13     $\pm$  0.24      &  1.83    $\pm$  0.15    &       1.72  $\pm$  0.29   &      1.49  $\pm$  0.14    \\
RR242\_28727   &  1.96     $\pm$ 0.11      &  3.66     $\pm$  0.35    &       2.06  $\pm$  0.29    &     1.08  $\pm$ 0.14   \\
RR242\_36267   &   2.72 $\pm$  0.28  &  0.37 $\pm$ 0.31 &  3.26  $\pm$ 0.41  &   1.62 $\pm$ 0.34 \\ 

\hline
\end{tabular}
}
\end{table}

Since our spectra do not extend up to the H$\alpha$ wavelengths, 
the emission correction to the H$\beta$ index was performed via the [OIII]$\lambda$5007
line. This is not optimal because the H$\beta$/[OIII] ratio as measured in samples of ETGs 
presents a large scatter: Trager et al.~(\cite{Tra00}) found that the H$\beta/$[OIII] ratio varies from 0.33 to 1.25,
with a median value of 0.6, while from the results of Annibali et al.~(\cite{Ann10}) we derive (excluding 
the Seyferts from the sample) a median value of  0.75 with a dispersion of 0.37.
We compute the emission correction to 
H$\beta$  to be ${\rm \Delta EW_{H\beta} = (0.75 \pm 0.37)  EW_{[OIII]}}$.
The [OIII] emissions, which are listed in Table~\ref{o3em},  
were derived subtracting simple stellar population models 
to the galaxy observed spectra, and fitting the emission lines in
the residual spectra with Gaussians of variable width, as fully described in 
Annibali et al.~(\cite{Ann10}).
Only for two galaxies, RR210\_11372 and RR242\_8064, a satisfactory fit could 
not be derived. 
Goudfrooij \& Emsellem~(\cite{goem96}) provide a correction to the Mgb index from the contamination of the [NI] emission line
doublet at 5199 \AA. The [NI]5199 is very faint compared to the other lines, and difficult to measure 
in our spectra. In the two galaxies with the strongest emission lines, RR242\_23187 and  RR143\_24246, we
derive ${\rm EW_{[NI]}/EW_{[OIII]} \sim}$ 0.01 and 0.03 respectively. 
This is significantly lower than the median value of $\sim$0.2 obtained from Table~1 of 
Goudfrooij \& Emsellem~(\cite{goem96}), assuming the same continuum at 5007 \AA~ and 5199 \AA, 
and could reflect the fact that dwarfs have lower metallicities than giant ellipticals.
According to our results, the contamination from the [NI]5199 doublet to 
the Mgb index is negligible.

Finally, we calibrated our indices into the Lick-IDS system through Lick standard stars, following the prescription 
by Worthey \& Ottaviani~(\cite{Wor97}). Five Lick stars  (HD~050778, HD~073665, HD~073710, HD~122563, and HD~136028)
 were observed together with the galaxies. The stellar spectra were degraded to match the Lick resolution, 
and the indices were measured following the same procedure as for the galaxies.
Then, our measurements were compared with the values provided by 
Worthey et al.~(\cite{worth94}) for the standard stars, and a linear transformation 
in the form ${\rm EW_{Lick}= \alpha \times EW_{us} + \beta}$ was derived through a least square fit.
This transformation was used to calibrate all the galaxy indices into the Lick-IDS system.

The errors on the indices were determined through the following procedure.
Starting from each real galaxy spectrum, we generated 1000 random  modifications  
by adding a wavelength dependent Poissonian fluctuation from the corresponding 
spectral noise $\sigma(\lambda)$. For each ``perturbed'' spectrum, we then repeated the index 
computation procedure, and derived the standard deviation.
The rms of the linear transformation into the Lick system was added in 
quadrature to the Poissonian error; for the H$\beta$ index, we added also the 
error associated with the emission correction.

We provide in Table~\ref{indices} the most important indices 
(H$\beta$, Mgb, Fe5270, Fe5335) for the individual galaxies.
These indices will be used in Section~3.2 to infer the stellar population parameters.
We excluded the galaxies  with very strong emission lines (RR143\_24246 and RR242\_23187), and with a very noisy spectrum  
(RR210\_11372). We end up with a sample of 14 galaxies. For RR242\_13326,  we do not provide the Fe5335 index,  
which is contaminated by a residual sky subtraction feature.

\begin{table}
\caption{Derived stellar population parameters.\label{tza}}
\begin{scriptsize}
\begin{tabular}{lccccc}
\hline \hline
Galaxy &  Log Age [yr]  & Log Z/Z$_{\odot}$\tablefootmark{a}  &  [$\alpha$/Fe]  &  [Fe/H]  &   Quality \tablefootmark{b}   \\   
\hline
 RR143\_09192   &     9.72 $\pm$0.20  &   -1.21$\pm$0.33   &   -0.80$\pm$0.16  & -0.59$\pm$0.37  & 2  \\   
RR210\_13493   &     9.76$\pm$0.17   &      0.02$\pm$0.23   &   -0.07 $\pm$ 0.12 & 0.08$\pm$0.26 & 1  \\  
RR216\_03519   &     9.92$\pm$0.34   &     -0.10$\pm$0.72   &   -0.47$\pm$0.19   & 0.29$\pm$0.74 & 1 \\  
RR216\_04052   &     9.75$\pm$0.16   &     -0.15$\pm$0.20   &   -0.04$\pm$0.15  & -0.11$\pm$0.25 & 1 \\  
RR216\_12209   &     9.58$\pm$0.26   &      -0.56$\pm$0.29  &   -0.69$\pm$0.17  & -0.01$\pm$0.34 & 1 \\  
RR242\_08064   &     9.57$\pm$0.14   &      -0.51$\pm$0.21   &  -0.23$\pm$0.20  & -0.31$\pm$0.29 & 1 \\  
RR242\_13326   &    $<$9.18  & $ >$ 0.08 & $<$ 0.8  & &  \\
RR242\_15689   &     9.83$\pm$0.18   &      -0.26 $\pm$0.21   &   -0.38 $\pm$0.14 &  0.06$\pm$0.25 & 1 \\  
RR242\_20075   &     9.91$\pm$0.18   &      -0.45 $\pm$0.24   &   -0.03 $\pm$ 0.20 & -0.42$\pm$0.31 &  1  \\ 
RR242\_22327   &   10.18$\pm$0.20  &       -0.10 $\pm$0.22   &    0.19 $\pm$0.14  & -0.27$\pm$0.26 & 1 \\
RR242\_24352  &      9.52$\pm$0.16   &       0.42 $\pm$0.29   &    0.03 $\pm$ 0.09  &  0.39$\pm$0.30 &  1 \\ 
RR242\_25575   &    9.18 $\pm$0.05   &      -0.42  $\pm$0.09  &    0.24 $\pm$ 0.19  & -0.64$\pm$0.21  & 1 \\ 
RR242\_28727   &   10.14$\pm$0.15   &      -0.45  $\pm$0.20  &   0.76 $\pm$0.19   &  -1.17$\pm$0.27 & 1 \\ 
RR242\_36267  &      9.30$\pm$0.14    &     -0.30  $\pm$0.23  &   -0.80 $\pm$ 0.12  & 0.31$\pm$0.26 & 2 \\
\hline
\end{tabular}
\tablefoot{
\tablefoottext{a}{Z$_{\odot}=0.0156$ from Caffau et al.~(\cite{caffau09}).}
\tablefoottext{b}{The quality of the fit is the best for flag=1 and the worst for 
flag=3 (see Section 3.2 for details).}
}
\end{scriptsize}
\end{table}

\subsection{Ages, metallicities and $\alpha$/Fe ratios}

To derive the stellar population parameters, we used the simple stellar population (SSP) models 
presented in Annibali et al.~(\cite{Ann07}), (hereafter A07). 
Here we extended the models down to ${\rm [\alpha/Fe]=-0.8}$  to account for the 
chemical compositions of dwarf galaxies\footnote{The SSPs are downloadable at {\tt http://web.oapd.inaf.it/rampazzo/SSPs-tables.html.}}.
We briefly recall that the SSPs are based on the Padova library of stellar 
models (Bressan et al.~\cite{Bre94}) and accompanying isochrones (Bertelli et al.~\cite{Ber94}).
The computation of the synthetic indices is
 based on the fitting functions (FFs) of Worthey et al.~(\cite{worth94}),
 and of Worthey \& Ottaviani~(\cite{Wor97}) for the higher order Balmer lines.
The departure from solar-scaled compositions is accounted for through 
the index responses of Korn et al.~(\cite{Kor05}).
As fully described in A07, we assign N, O, Ne, Na, Mg, Si, S, Ca and Ti 
to the {\it $\alpha$ group}, and Cr, Mn, Fe, Co, Ni, Cu and Zn to the {\it Fe-peak group}.
For simplicity, we assume in these models that the elements are enhanced/depressed by the same factor within each group,
even if it has been shown that they exhibit different trends with [Fe/H] (e.g., Fulbright et al.~\cite{ful07}).
While the  $\alpha$ and  Fe-peak groups are
respectively enhanced and depressed  in the [$\alpha$/Fe]$>0$ models, the opposite holds 
for the [$\alpha$/Fe]$<0$ models. 
The effect of departure from solar-scaled compositions on the stellar evolutionary tracks is not included 
in these models.

As shown by several authors (e.g., Worthey et al.~\cite{worth94}; Thomas et al.~\cite{Tho03}; A07) a 
three-dimensional space defined by a Balmer absorption line (H$\beta$, H$\gamma$, or H$\delta$),
the metallicity sensitive [MgFe]$^{'}$ index (defined as ${\rm \sqrt{ Mgb \times (0.72 \ Fe5270 + 0.28 \ Fe5335)}}$),
and a metallic index particularly sensitive to the $\alpha$-elements (like Mgb), turns out a powerful 
diagnostic for the derivation of ages, metallicities, and [$\alpha$/Fe] ratios in unresolved stellar systems. 
The H$\beta$ index is poorly affected by $\alpha$/Fe variations, but presents the disadvantage that can be 
significantly contaminated by nebular emission. The  bluer H$\gamma$ and  H$\delta$ lines are a valuable alternative to H$\beta$ because they are less affected by emission. However, because of the limited 
wavelength range covered by our spectra, these lines are measured only in few galaxies. 
To avoid systematic effects caused by the use of H$\gamma$ and  H$\delta$ only in some galaxies, 
we decided to base the entire derivation of the stellar population parameters uniquely on H$\beta$.

\begin{figure*}
  \centering
  \includegraphics[width=19cm]{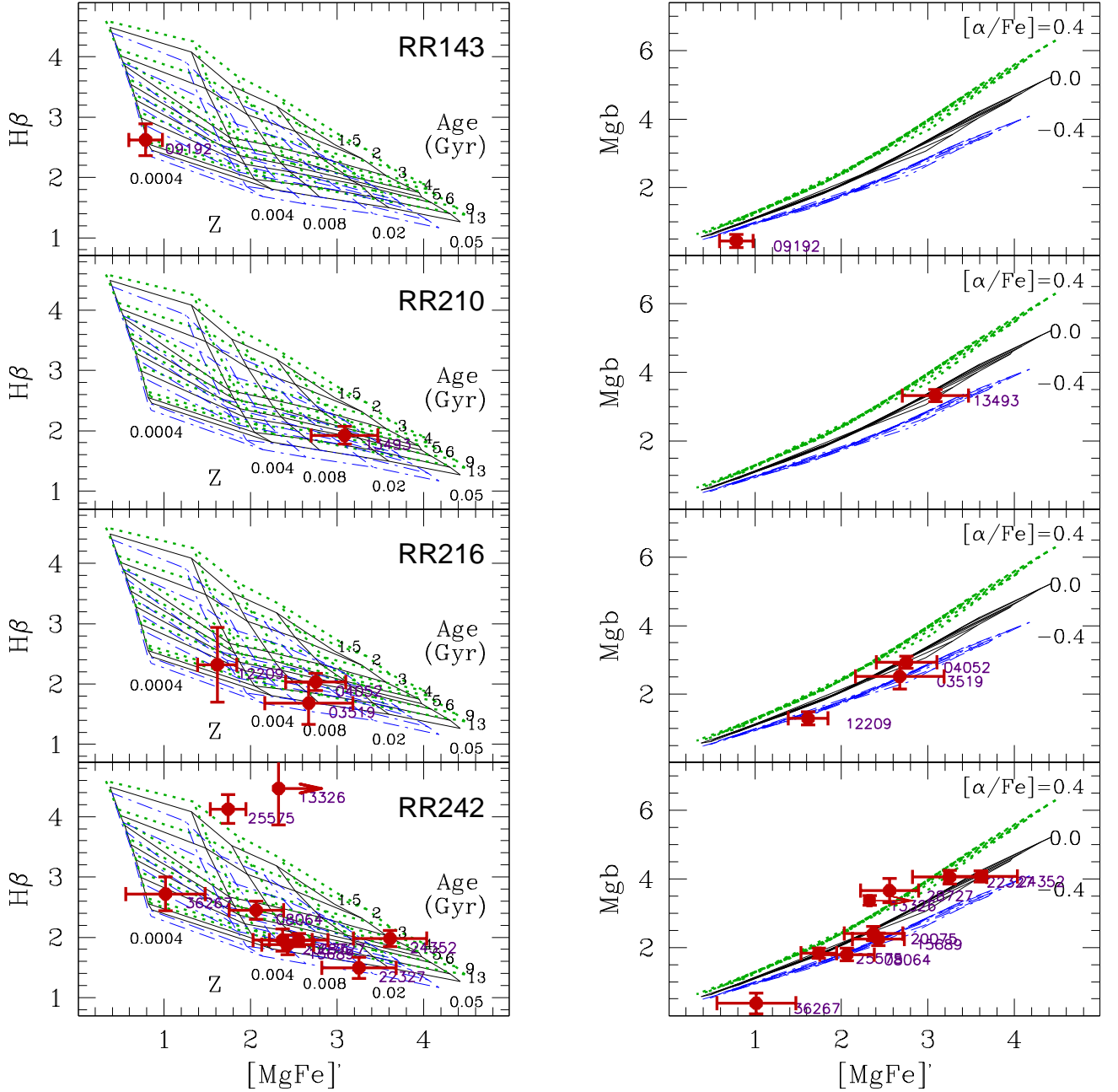}
  \caption{Lick indices (H$\beta$ vs.  [MgFe]$^{'}$, left panels, and Mgb vs. [MgFe]$^{'}$, right panels) 
  extracted at  ${\rm r<r_e/2}$ for the dwarf sample.  
  From top to bottom, the galaxies were separated according to their group membership:
  RR143, RR210, RR216, and RR242.  For RR242\_13326,  [MgFe]$^{'}$  is a lower limit.
  The data are compared with SSP models of metallicities Z$=$0.0004, 0.004, 0.008, 0.02, 0.05, ages$=$ 13, 9, 6, 5, 4, 3, 2, 1.5 Gyr, and 
  [$\alpha$/Fe]$=$-0.4, 0., 0.4. }
\label{fig1}
\end{figure*}

Two projections of the 3D H$\beta$$-$[MgFe]$^{'}$$-$Mgb space 
are shown in Fig.~\ref{fig1}. 
In the H$\beta$ vs. [MgFe]$^{'}$ plane, models of constant age run almost horizontal, while models 
of constant metallicity run almost vertical. This plane is poorly affected by $\alpha$/Fe variations, and is 
well suited to directly reading off ages and total metallicities almost independently 
of the specific chemical composition.
In this plane, our galaxies cover the largest possible range in age ($\sim$ 1-13 Gyr) and 
metallicity (Z$=$0.0004-0.05).

In the Mgb vs. [MgFe]$^{'}$ plane, models of constant age and metallicity are almost degenerate, 
while the [$\alpha$/Fe] ratio is well separated. 
We notice that the bulk of our data is consistent with solar or subsolar [$\alpha$/Fe] ratios
(down to $-$0.4 and even less), indicating an enhancement of the Fe-peak elements over the $\alpha$-elements compared to the solar composition.

Ages, metallicities, and  [$\alpha$/Fe] ratios were derived  through the algorithm described in A07.
Each model of given age (t), metallicity (Z), and [$\alpha$/Fe] ratio ($\alpha$) univocally corresponds to a point in the 
H$\beta$-$<$Fe$>$\footnote{$<Fe>=\frac{1}{2}(Fe5270 + Fe5335)$}-Mgb three-dimensional space.  Given an observed index triplet 
(H$\beta_o$ $\pm$ $\sigma_{H\beta}$, $<$Fe$>_o$  $\pm$ $\sigma_{<Fe>}$, Mgb$_o$ $\pm \sigma_{Mgb}$), 
it is therefore possible to compute the probability $P_{t, Z, \alpha}$ that the 
generic (t, Z, $\alpha$) model is the solution to that data point:

\begin{eqnarray}
P_{t,Z,\alpha}&=& \frac{1}{\sqrt{(2 \pi)^3} \sigma_{H\beta}  \sigma_{<Fe>}  \sigma_{Mgb}} 
\exp \left[ -\frac{1}{2} \left( \frac{H\beta-H\beta_o}{\sigma_{H\beta}} \right)^2 \right.   \nonumber \\
 &-&\left. \frac{1}{2} \left( 
\frac{<Fe>-<Fe>_o}{\sigma_{<Fe>}}\right)^2  -\frac{1}{2} \left(\frac{Mgb-Mgb_o}{\sigma_{Mgb}}\right)^2 \right]
\end{eqnarray}

\noindent where (H$\beta$, $<$Fe$>$, Mgb) is the index triplet associated to the (t, Z, $\alpha$) model.
Eq. (1) holds if the errors are Gaussians. 
This allows us to derive a probability density map $P_{t, Z, \alpha}$ defined on the 3D  
(t, Z, $\alpha$) space. 
Then the solutions are computed to be:  

\begin{eqnarray}
X_{\mu}= \frac{\int\!\!\!\int\!\!\!\int_{G^*} X \  P(t,Z,\alpha) \ dt \ dZ \ d\alpha}{\int\!\!\!\int\!\!\!\int_{G^*}  P(t,Z,\alpha) \ dt \ dZ \ d\alpha} \nonumber \\
\end{eqnarray}

\noindent where X is the age t, the metallicity Z, or the composition $\alpha$, and $G^*$ is the subspace of solutions 
defined by $P(t,Z,\alpha) > 0.95 P_{max}$.
The errors on the solutions are:

\begin{eqnarray}
{\sigma_X}^2 &=& \frac{\int\!\!\!\int\!\!\!\int (X-X_{\mu})^2 \  P(t,Z,\alpha) \ dt \ dZ \ d\alpha}{\int\!\!\!\int\!\!\!\int  P(t,Z,\alpha) \ dt \ dZ \ d\alpha} \nonumber \\
\end{eqnarray}

\noindent computed on the whole (t, Z, $\alpha$) space.

The goodness of the solution is evaluated  through  the ``distance''  between the data and the solution.
From $P_{best fit}$, the probability obtained  substituting into equation (1) the index values of the best fit model, we define three quality flags:

\begin{equation}
C \times P_{best fit} \ge \exp \left[-3/2 \right]  \ \ \ flag=1 
\end{equation}
\begin{equation}
\exp \left[-6 \right] \le C  \times P_{best fit} < \exp \left[-3/2 \right]  \ \ \ flag=2 
\end{equation}
\begin{equation}
C \times P_{best fit} < \exp \left[-6 \right]  \ \ \ flag=3 
\end{equation}

\noindent where $C= \sqrt{(2 \pi)^3} \sigma_{H\beta}  \sigma_{<Fe>}  \sigma_{Mgb}$. 
The adopted limits for the three flags correspond respectively to 1 $\sigma$, 2 $\sigma$, and 3 $\sigma$ difference between 
the observed indices and the best-fit model; thus the quality of the solution is the best for flag$=1$, and the worst for flag$=3$.
The derived stellar population parameters, as well as the quality of the solutions,  are given in Table~\ref{tza}.
Notice that, because of the indices used to derive the stellar population parameters, [$\alpha$/Fe] directly reflects a [Mg/Fe] ratio.
Because of the lack of the Fe5335 index, only upper/lower limits could be derived 
for RR242\_13326. Two galaxies, namely RR143\_09192 and RR242\_36267, 
have a quality flag of 2. They are very low in Mgb, and our algorithm provides for them 
the lowest [$\alpha$/Fe] values allowed by the models. Likely, the errors on their indices were underestimated.
For all the other galaxies, the quality flag is 1.

Concluding, we derive the stellar population parameters for 13 dwarf galaxies (plus one upper/lower limit).
The median values and scatters around the medians for age, $\log Z/Z_{\odot}$ and [$\alpha$/Fe] are 
$5.2\pm4.1$ Gyr, $-0.42\pm0.38$, and $-0.23\pm0.44$, respectively.
For the subsample of 12 dwarf early-types, the medians are: $5.7\pm4.2$ Gyr, $-0.30\pm0.37$, and $-0.07\pm0.43$.
If we exclude the two quality-flag$=$2 objects (RR143\_09192 and RR242\_36267) we obtain, for
a fiducial subsample of 10 early-type dwarfs, values of  $5.7 \pm4.4$ Gyr, $-0.26\pm0.28$, and  $-0.04\pm0.33$.

\section{Correlations with internal galaxy properties: nature}\label{internal}

 \begin{figure}[h!]
  \centering 
  \includegraphics[width=9cm]{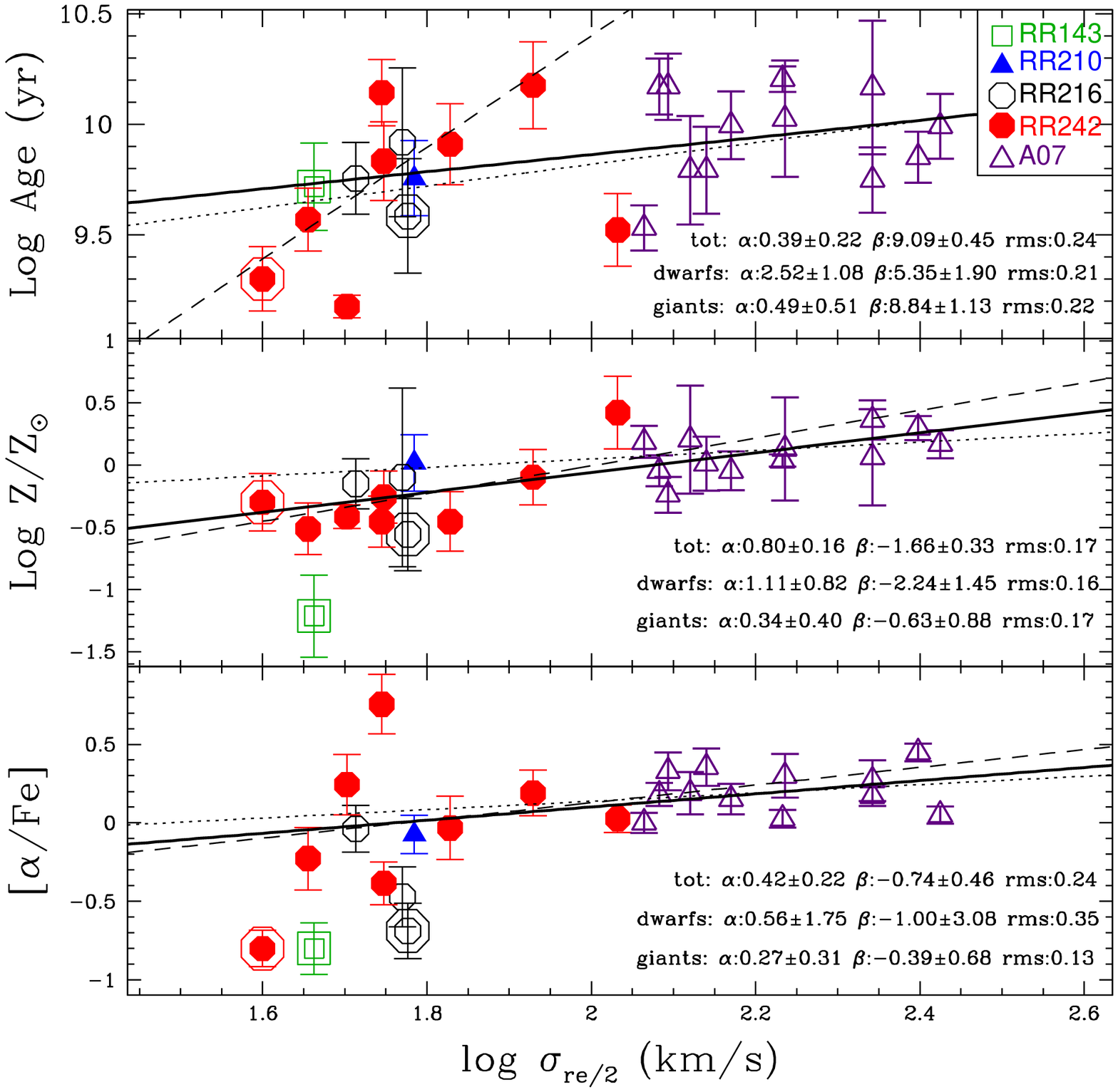}
  \caption{
  Ages, metallicities, and [$\alpha$/Fe] ratios at  ${\rm r<r_e/2}$ versus  
  velocity dispersion.  Different symbols indicate the group membership.
For comparison, we plot also the stellar population parameters for a selection of 
galaxies from the bright ETGs of A07 (open triangles, see text for details).
The solid line indicates the least square linear fit to the (dwarf $+$ giant) ETG sample, 
where we have excluded the Sc galaxy RR216\_12209 (big open circle), and the 
two low quality-fit galaxies (see Sect~3.2 for details): RR143\_09192 (big open square) 
and RR242\_36267 (big filled circle).
The  dashed and dotted lines are the separate fits to the dwarf ($\sigma_{r_e/2}<$100 km s$^{-1}$) 
and giant  ($\sigma_{r_e/2}\ge$100 km s$^{-1}$) samples, respectively. The slope ($\alpha$), zero-point ($\beta$) and 
scatter of the fits are indicated in the panels. }
\label{sigma}
\end{figure}

\subsection{Correlations with velocity dispersion}

Studies of integrated stellar populations in ETGs have revealed the presence 
of scaling relations between the luminosity weighted age, metallicity, or [$\alpha$/Fe] ratio, and the galaxy velocity dispersion (e.g., Thomas et al.~\cite{Tho05}; Clemens et al.~\cite{Cle06}; A07; Clemens et al.~\cite{Cle09}).
These relations testify that the galaxy potential well is the main driver of the evolution of the stellar populations.
The average stellar metallicity and the [$\alpha$/Fe] ratio increase with $\sigma$, indicating that the chemical enrichment was 
more efficient and faster in more massive galaxies. At the same time, an increase in galaxy ages with $\sigma$ is observed. 
These scaling relations suggest  ``downsizing'' in galaxy formation, meaning that 
more massive galaxies formed their stars earlier and faster than lower mass galaxies.

The galaxy sample examined in this paper covers velocity dispersions (measured at 
${\rm r<r_e/2}$) in the range 40 --108 km s$^{-1}$ and absolute R magnitudes in the range $-14.8$ -- $-19.8$ mag (see Table~\ref{dwarf_properties}).
This allows us to extend the scaling relations usually derived for giant ETGs to the low-$\sigma$ regime.
We plot the derived ages, metallicities, and [$\alpha$/Fe] ratios versus $\sigma_{r_e/2}$ in Fig.~\ref{sigma}.
For comparison, we include a selection of galaxies  from the bright ETGs of A07 satisfying the following requirements:
1) reside in comparable environment as our dwarfs according to the volume density (see Col.~8 of Table~\ref{dwarf_properties}), i.e.,  $\rho_{xyz} \lesssim 1.2 Mpc^{-3}$ (Tully~\cite{Tul88});
2) have a measure of $\sigma_{r_e/2}$; 3) have absent or very low nebular emission according to A10.
The last requirement is due to the fact that, for a self-consistent comparison between the two samples, 
the stellar population parameters  for the A07 galaxies must be derived using the same indices (H$\beta$, [MgFe]$^{'}$, Mgb), SSPs, and algorithm as for the dwarf sample  (they can be retrieved from the electronic Table~5 of A07).
Because of the strong uncertainly in the H$\beta$ emission correction, and the impossibility of 
using the higher order Balmer lines, we have to exclude the A07 galaxies with strong emission lines.
We end up with 12 galaxies from the A07 sample (NGC~1407, NGC~1426, NGC~3557, NGC~3818, NGC~4697, NGC~5638, NGC~5812, NGC~5813, NGC~5831, NGC~6721, NGC~7192, and NGC~7332), which have $\sigma_{r_e/2}$ in the range 120 -- 270 km s$^{-1}$,
and absolute R magnitudes in the range $-20.8$ -- $-22.9$.
In Fig.~\ref{sigma}, we show the least square fits to the total dwarf$+$giant ETG sample. 
We have excluded from the fit the Sc galaxy RR216\_12209, and the two galaxies 
 (RR143\_09192 and RR242\_36267) for which the quality of the SSP fit was not 
 very good (quality flag$=2$, see Section~3.2 for details). 
 We derive the following relations:

\begin{eqnarray}
\log Age[yr]&=& (0.39\pm0.22) \times \log \sigma_{r_e/2} [km s^{-1}] + ( 9.09\pm0.45) 
\nonumber \\ 
&&~~~~~~~~~~~~~~~~~~~~~~~~~~~~~~~~~~~~~~~~~~~~~[rms=0.24]
\end{eqnarray}

\begin{eqnarray}
\log Z/Z_{\odot}&=& (0.80\pm0.16)  \times \log \sigma_{r_e/2} [km s^{-1}] -(1.66\pm0.33) 
\nonumber \\ 
&&~~~~~~~~~~~~~~~~~~~~~~~~~~~~~~~~~~~~~~~~~~~~~~~~~[rms=0.17]
\end{eqnarray}

\begin{eqnarray}
[\alpha/Fe]&=& (0.42\pm0.22)   \times \log \sigma_{r_e/2} [km s^{-1}] -(0.74\pm0.46) 
\nonumber \\ 
&&~~~~~~~~~~~~~~~~~~~~~~~~~~~~~~~~~~~~~~~~~~~~~~~~~~~~[rms=0.24]
\end{eqnarray}

Eqs. (7) through (9) show that age, metallicity, and [$\alpha$/Fe] ratio tend to increase with $\sigma$.
Lower mass galaxies tend to be younger, metal poorer, and less enhanced in the 
$\alpha$-elements (with solar or even subsolar [$\alpha$/Fe] ratios) than more massive ETGs.
This is consistent with the extrapolation of the trends seen in bright ETGs.
A (non parametric) Spearman rank-order test gives correlation coefficients of 0.37, 0.73, and 0.45 (N galaxies$=$22) for 
age, metallicity, and [$\alpha$/Fe] ratio, respectively, corresponding to probabilities of  91\%,  99.99 \%, and 96.44\% for 
correlations to exist.  Thus the strongest correlation is between metallicity and $\sigma$, while age is weakly correlated. 
This is in agreement with previous studies finding strong correlations between metallicity or [$\alpha$/Fe]
and $\sigma$, but weaker correlations between age and $\sigma$ (e.g., Kuntschner et al~\cite{kunt01}; 
Thomas et al.~\cite{Tho05}).

In Fig.~\ref{sigma} we also provide the parameters of the separate fits to the 
 dwarf ($\sigma_{r_e/2}<$100 km s$^{-1}$) and to the giant  
($\sigma_{r_e/2}\ge$100 km s$^{-1}$) samples.
RR242\_24352, with $\sigma \sim$108 km~s$^{-1}$, is fitted together with the giants. 
This galaxy is on the upper end of the faint galaxy range and at the lower end 
of the giant range of our sample (see the Hamabe-Kormendy relation in Fig.~11 of Gr\"utzbauch et al.~(\cite{Gru09})).
The trends are consistent with the relations derived for the total 
(dwarf$+$giant) sample. Notice however the large errors on the fit coefficients, especially for the [$\alpha$/Fe] vs. $\sigma$ relation of dwarfs.
The slope of the age versus $\sigma$ 
relation is much steeper for the dwarfs than for the giants. 
This is in agreement with Nelan et al.~(\cite{nel05}), finding a steepening of the slope of the age vs. $\sigma$ relation 
at the low mass regime. 
The correlation coefficients for the separate fits to dwarfs (9) and giants (13) are 0.73 and 0.25, 0.6 and 0.13, $\sim$0 and 0.33 for 
age, metallicity, and [$\alpha$/Fe] ratio, respectively.

\subsection{Correlations with morphology}

 \begin{figure*}
  \centering
  \includegraphics[width=0.485\textwidth]{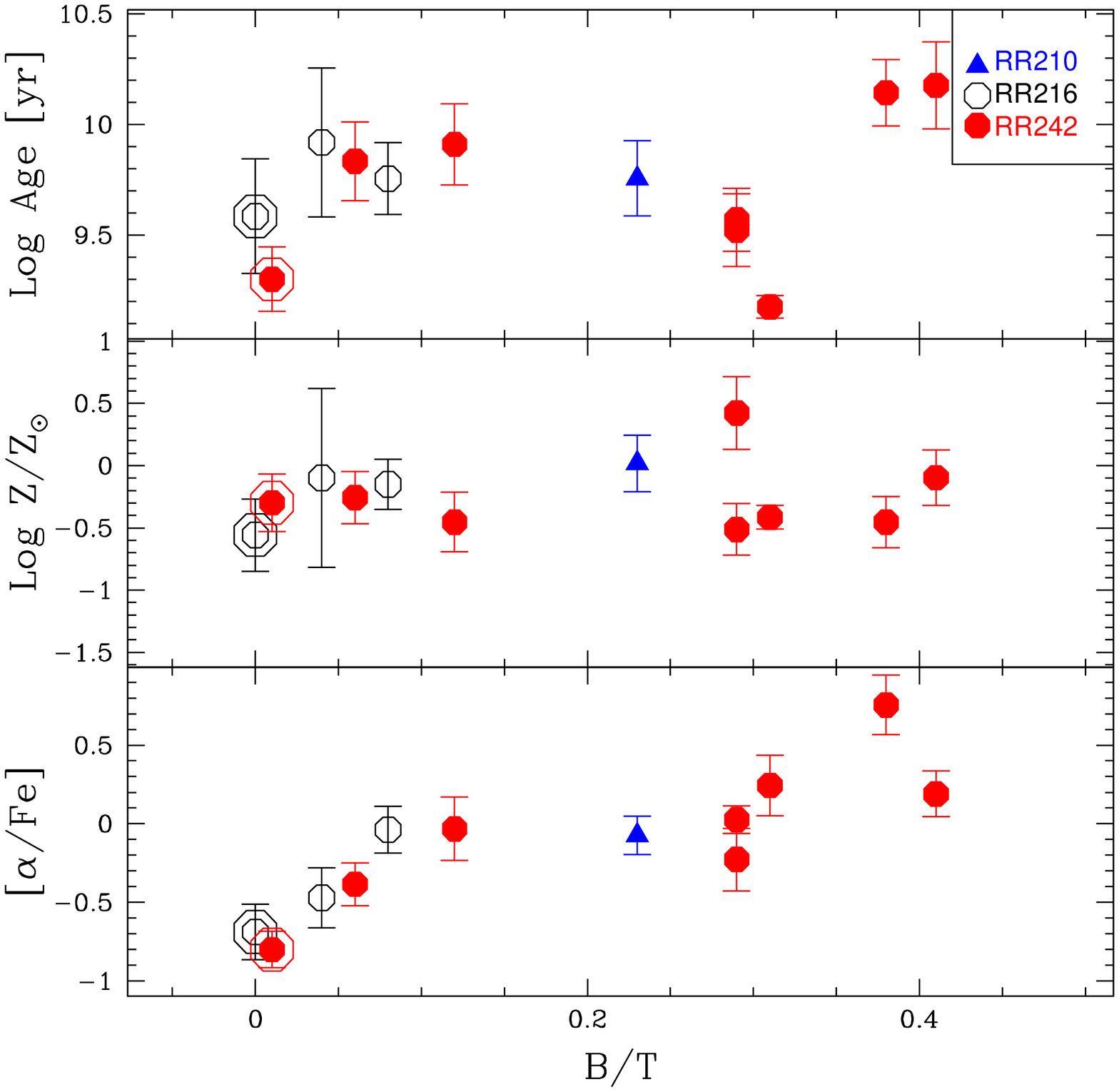}
  \includegraphics[width=0.485\textwidth]{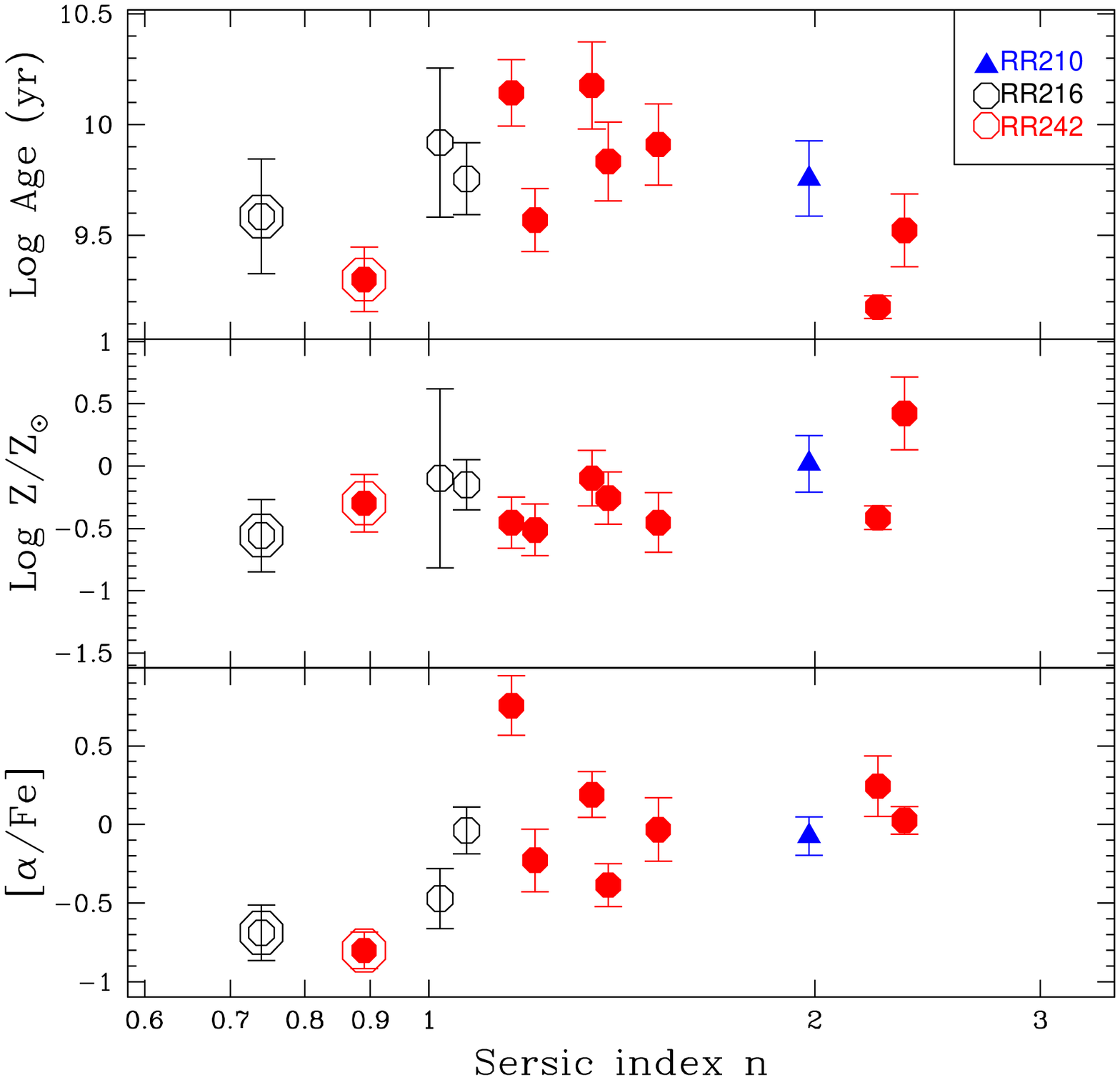}
  \caption{Stellar population parameters (ages, metallicities, and [$\alpha$/Fe] ratios) as a function of morphology. Left panel: bulge fraction B/T, right panel: Sersic index $n$. Different symbols indicate the group membership: RR~210 (blue triangles), RR~216 (black open circles) and RR~242 (red full dots). We remark with larger symbols the Sc galaxy RR216\_12209 (open circle) and the low quality fit galaxy 
 RR242\_36267 (filled circle)}.
\label{morph_indices}
\end{figure*}

In this section we investigate possible relations between the stellar population parameters and the galaxy's morphology, i.e. 
the bulge fraction, B/T, and the Sersic index $n$.  
Figure~\ref{morph_indices} shows ages, metallicities, and [$\alpha$/Fe] ratios as a function of  B/T (left panel) and  $n$ (right panel)  
for our  sample. 
In the plot, we do not show RR143\_09192, since its morphological parameters are not robust due to a bright star very close to the centre of this galaxy. We will also exclude from the following analysis RR242\_36267, which has a quality flag$=2$ in the derived stellar population parameters (see Sect.~3.2), and  RR216\_12209, which is classified as an Sc. 
A (non parametric) Spearman rank-order test gives correlation coefficients of 0.04, $-0.05$, and 0.83 with 
N galaxies$=$10  for age, metallicity, and [$\alpha$/Fe] ratio versus B/T, respectively, indicating that the only strong correlation is 
between [$\alpha$/Fe] and the bulge-to-total light fraction (P$\sim$ 99.7 \%).
 For the Sersic index $n$ we derive lower correlation coefficients of $-$0.53, 0.31, and 0.33. 
We notice  that there is one outlier, RR242\_ 28727, which has a high B/T ratio but a relatively small $n$. This galaxy is located very close to the bright elliptical of the pair and the Sersic fit might be biased by the light of the close bright neighbour.

\begin{figure}
  \centering
  \includegraphics[angle=0, width=0.485\textwidth]{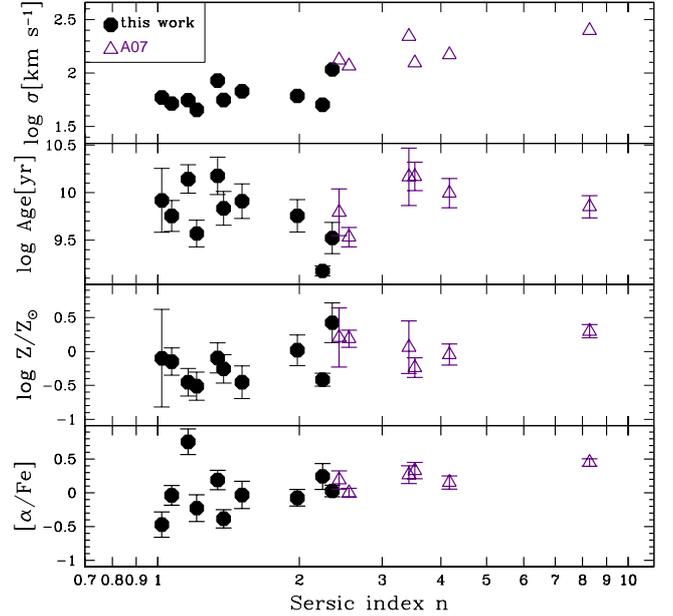}
  \caption{Velocity dispersion $\sigma$, age, metallicity and [$\alpha$/Fe] ratio as a function of the 
  Sersic index $n$. Full dots are the early-type dwarfs of this paper, while open triangles 
  are the bright ETGs selected from A07 (see text for details).
 We excluded RR143\_09192 and RR242\_36267, for which we obtain a low quality fit (see Sect.~3.2),
  and  RR216\_12209, which is classified as an Sc. }
\label{n_afe_bright}
\end{figure}

 In Figure~\ref{n_afe_bright}, we study the relationship between stellar population parameters versus morphology on a larger luminosity baseline, by adding the bright ETGs of A07.
Sersic indices were measured from SDSS images for 6 galaxies among the subset of the A07 sample
(Marino et al.~\cite{marino10}).
Bulge-to-total light fractions for the A07 sample are not available. 
The Spearman correlation coefficients are 0.07, 0.58, and 0.54 (N$=$16) for age, metallicity, and [$\alpha$/Fe] versus $n$, respectively, indicating that significant correlations are present only for metallicity and  [$\alpha$/Fe] (with probabilities of 98.1\% and 96.9\%, respectively).
 The presence of correlations between metallicity and $n$, and between  [$\alpha$/Fe] and $n$, is expected 
 from the fact that the metallicity and the [$\alpha$/Fe] ratio correlate with $\sigma$, 
 and $\sigma$ correlates with $n$ (see top panel of Figure~\ref{n_afe_bright}).

\section{The role of environment: nurture}\label{environment}

\subsection{Correlations with local galaxy density}

\begin{figure}
  \centering
  \includegraphics[width=0.485\textwidth]{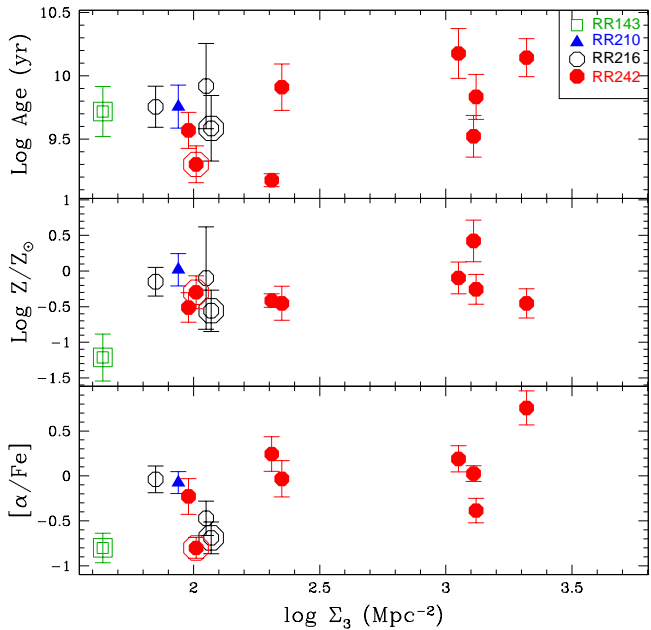}
  \caption{Stellar population parameters as a function of surface local density. From top to bottom: age, metallicity and [$\alpha$/Fe] ratio. The symbols are the same as in Figure~\ref{sigma}.}
\label{dens_indices}
\end{figure}

To investigate the effect of local galaxy density within the group environment, we plotted 
the stellar population parameters versus  $\Sigma_3$ in Figure~\ref{dens_indices}.
As already explained in Section~ 2.5, $\Sigma_3$ is sensitive to very local galaxy concentrations, 
and can vary significantly within the same group. 

If we exclude RR143\_09192 and RR242\_36267, for which we obtain a low quality fit (see Sect.~3.2), 
and  RR216\_12209, which is classified as an Sc, we derive Spearman correlation coefficients of 0.35, $-0.1$, and 
0.42 (10 data points) for age, metallicity, and  [$\alpha$/Fe] ratio, 
indicating that there is no significant correlation. 
However,  $\Sigma_3$ is a {\it projected} galaxy density, 
and projection effects can considerably distort the intrinsic 3D galaxy density.
A de-projected density could be derived assuming a certain distribution for the galaxies in the 3D space (e.g., Poggianti et al.~\cite{pog10}), but  due to the low number of galaxies in each group, a statistical correction based on the assumption that the galaxies are evenly distributed within the group halo is not very meaningful.
Notice also that $\Sigma_3$ represents an upper limit to the 3D galaxy density.

\subsection{Comparison with the cluster environment}

According to the large-scale densities $\rho_{xyz}$ given in Col.~8 of Table~\ref{dwarf_properties}, our dwarf galaxies are representative of the low-density environment.
In this Section, we compare the properties of our sample with those of cluster dwarfs previously studied in the literature. 

Line-strength index studies of dwarf early-type galaxies were mainly preformed in the Virgo 
(Gorgas et al~\cite{gorg97}; Geha et al.~\cite{geha03}; van Zee et al.~\cite{vanz04}; 
Michielsen et al.~\cite{mich08}; Paudel et al.~\cite{pau10}) and Coma (Poggianti et al.~\cite{pog01a}; 
Smith et al.~\cite{smith09}; Matkovi{\'c} et al.~\cite{matk09}) clusters. 
However, when comparing our dataset with others, it is important to make sure that 
the extraction apertures sample comparable portions of the galaxy light, because of the presence 
of radial gradients in the stellar populations.
The relatively short distance to Virgo implies that literature data typically sample very central 
galaxy regions ($r<$0.15 $r_e$), and a comparison with our larger aperture data is not much meaningful.
On the other hand, literature data for Coma, at a larger distance than Virgo, typically sample larger apertures, and are more suitable for a comparison with our sample.
 Smith et al.~(\cite{smith09}) provide Lick indices for faint red galaxies 
observed in two 1$^o$ fields, one centered on the Coma cluster core and the other located 
 1$^o$ to the south-west of the cluster center. The galaxies have M$_r \lesssim -17$, and velocity dispersions 
 in the range 20-80 km$^{-1}$.
 The spectra were collected with a 1.5'' fiber diameter, corresponding, 
given the median effective radius of $r_e \sim 3''$, to an $r<r_e/4$ aperture. 
Since the use of a r$_e$/4 instead of a r$_e$/2 aperture 
implies only small offsets in $\delta$ log t, $\delta$ log Z, and $\delta$ [$\alpha$/Fe] 
($\sim$ 0.037, 0.025 and $<0.02$ at $\sigma \sim$ 100 km s$^{-1}$ from Fig.2 of  Clemens et al.~\cite{Cle09}), 
the Smith et al.~(\cite{smith09}) data are suitable for comparison with our sample. 
We select the Coma dwarfs in the 1$^o$ field centered on the cluster core. 

To avoid systematic effects due to the use of different stellar population models, we reprocessed the 
H$\beta$, Fe5270, Fe5335, and  Mgb indices provided by Smith et al.~(\cite{smith09}), their Table~E1,  
with our own SSPs according to the procedure described in Section~3 (the derived ages, metallicities, and [$\alpha$/Fe]
ratios are given in  Appendix~\ref{appendixa}).
In Fig.~\ref{comasigma} we show the Coma dwarfs in the Age/ Z/ [$\alpha$/Fe] versus $\sigma$ planes,
together with the dwarfs our sample (low density environment, LDE) and the giants of A07.  
We draw least square linear fits to the (Coma dwarfs$+$A07 giants) and to the (LDE dwarfs$+$A07 giants) samples separately. 
A comparison of the fit parameters provided in Fig.~\ref{comasigma} and Fig.~\ref{sigma} 
shows that the slopes are compatible within the errors.

 \begin{figure}
  \centering
  \includegraphics[width=9cm]{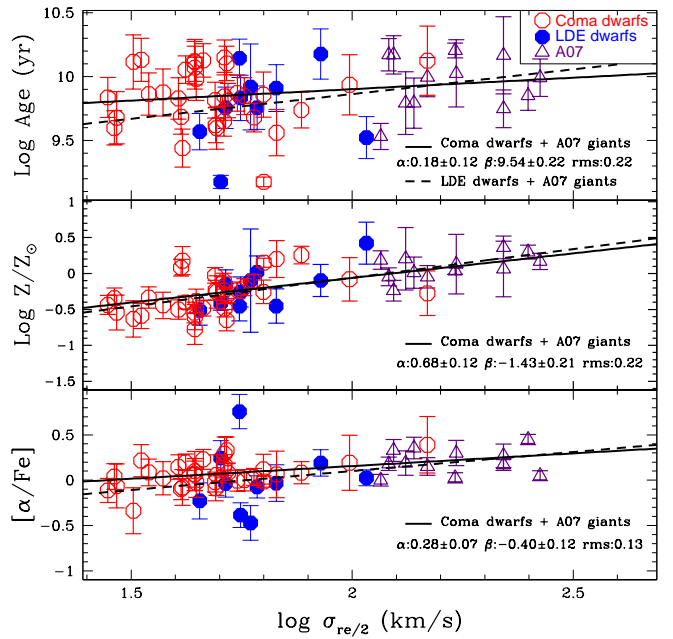}
  \caption{Age, metallicity, and [$\alpha$/Fe] ratio versus $\sigma$. 
  Open circles are the Coma dwarfs from Smith  et al.~(\cite{smith09}), filled circles are the low-density 
  environment (LDE) dwarfs of this paper (we have excluded RR143\_09192 and RR242\_36267, which have a poor quality SSP fit, see Sect.~3.2, and RR216\_12209, which is an Sc), while open triangles are the ETGs of A07. The solid and dotted lines are the least square linear fits to the (Coma dwarfs$+$A07) and (LDE dwarfs$+$A07) samples, respectively.  } 
\label{comasigma}
\end{figure}

 \begin{figure*}
  \centering
  \includegraphics[width=16cm]{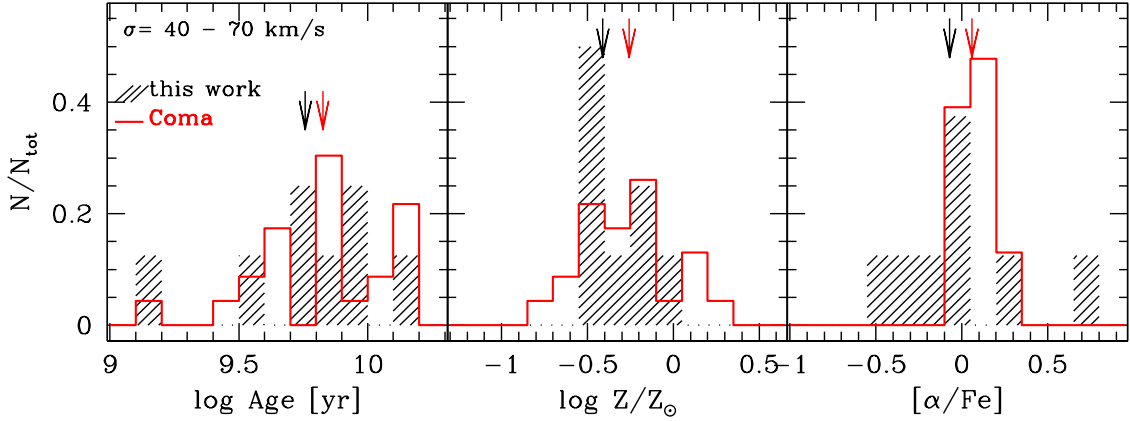}
  \caption{Age, metallicity, and [$\alpha$/Fe] distributions for our early-type dwarfs and for red passive dwarfs in a field 
  centered on the Coma core. All galaxies were selected to have $\sigma$ in the range 40-70 km s$^{-1}$.
  Vertical arrows indicate the median values for the two samples.
Within our sample, we did not include  RR143\_09192 and RR242\_36267, which have a poor quality SSP fit (see Sect.~3.2), 
and the Sc galaxy RR216\_12209}.
\label{coma}
\end{figure*}

To isolate the effect of environment, we select galaxies with velocity dispersions in the range 40-70 km s$^{-1}$, and 
investigate possible differences in age, metallicity, and [$\alpha$/Fe].
We do not include  RR143\_09192 and RR242\_36267, which have a poor quality SSP fit (see Sect.~3.2), 
and the Sc galaxy RR216\_12209.
The result of this analysis is shown in Figure~\ref{coma}, where we plot the distributions of the stellar population parameters 
for our early-type dwarfs and for the Coma red passive dwarfs.
The median ages, metallicities ($\log Z/Z_{\odot}$), and  [$\alpha$/Fe] ratios are
 5.7$\pm$3.6 Gyr and  6.7$\pm$3.9 Gyr,  $-0.41\pm0.22$ and $-0.26\pm0.26$, 
and $-0.07\pm0.36$ and  $0.06\pm0.11$ for us and Coma, respectively.
At face values, the medians indicate that low-density environment dwarfs are on average younger, less metal rich, and less 
enhanced in [$\alpha$/Fe] (longer star formation timescales) than Coma dwarfs.
We performed statistical tests to check the significance of the difference between the two samples. 
According to a Mann-Withney U test, the age, metallicity, and [$\alpha$/Fe] distributions shown in Fig.~~\ref{coma} are not significantly 
different. On the other hand, a two-sample Kolmogorov-Smirnov test provides $D_{n,m}$ values of 0.58, 0.51, and 1.61 for age, metallicity, and 
[$\alpha/Fe$], respectively, corresponding to probabilities of 0.89, 0.96, and 0.01 that the two datasets come from the same distribution. 
The K-S test implies that there is a difference in [$\alpha$/Fe] between the two samples, since the probability that they come from the same 
distribution is only $\sim$1\%. 
Concluding, we find evidence that low-density environment dwarfs experienced longer star formation timescales than their cluster counterparts. 
However, these results need to be confirmed through larger data samples.

\section{Discussion and conclusions}\label{discussion}

Using Lick indices, we have derived luminosity-weighted ages, metallicities, and [$\alpha$/Fe] ratios for 
a sample of 13 dwarf galaxies, 12 of which are early-type, located in poor groups dominated by a bright E+S 
galaxy pair (Gr\"utzbauch et al.~\cite{Gru09}).
The groups typically have less than 5 bright ($\sim L^\ast$) galaxies, 
and are characterized by local densities  (as defined by Tully~\cite{Tul88}) 
of $\rho_{xyz} \lesssim 1.2$. The dwarf galaxies in our sample therefore reside in low density environments.
They have velocity dispersions in the range (40 - 108) km s$^{-1}$ and 
M$_R$ varies from $-19.8$ to $-14.8$.

For our fiducial subsample of 10 dwarf early-type galaxies, we derive 
median values and scatters around the medians of $5.7 \pm4.4$ Gyr, $-0.26\pm0.28$, and  $-0.04\pm0.33$ 
for age, $\log Z/Z_{\odot}$, and $[\alpha/Fe]$, respectively. 
For a selection of bright ETGs from the A07 sample residing in comparable environment
we derive median values of $9.8 \pm4.1$ Gyr, $0.06\pm0.16$, and  $0.18\pm0.13$
for the same stellar population parameters.
Thus, it emerges that dwarfs  are on average younger, less metal rich, and less enhanced in the $\alpha$-elements than giants,  in agreement with other studies of dwarf galaxies (e.g., Geha et al.~\cite{geha03}; van Zee et al.~\cite{vanz04}; Michielsen et al.~\cite{mich08}; Smith et al.~\cite{smith09}, Spolaor et al.~\cite{spo10}), 
and with the extrapolation to the low-mass regime of the scaling relations 
derived for giant ETGs (e.g., Nelan et al.~\cite{nel05}, Thomas et al.~\cite{Tho05}; Clemens et al.~\cite{Cle09}).
Considering the total (dwarf$+$giant) sample, we derive that age $\propto \sigma^{0.39\pm0.22}$, $Z\propto \sigma^{0.80\pm0.16}$, and $\alpha$/Fe $\propto \sigma^{0.42\pm0.22}$.

The low [$\alpha$/Fe] ratios derived for the dwarfs imply long star formation (SF) timescales, 
as opposed to the short SF timescales in giant ETGs, in agreement with 
``downsizing'' in galaxy formation. 
In fact, $\alpha$-elements are mainly produced in massive stars and restored into the interstellar medium (ISM) through the explosion of Type II SNe on short SF timescales ($<$ 50 Myr). 
On the other hand, iron, mostly provided by Type Ia SNe, 
which are associated with the evolution of intermediate and low-mass stars, 
is injected into the ISM on much longer timescales: the time in maximum enrichment 
by SNIa varies from $\sim$40$-$50 Myr for an instantaneous starburst to  $\sim$0.3 Gyr for a typical 
elliptical galaxy to $\sim$4.0$-$5.0 Gyr for a spiral galaxy (Matteucci \& Recchi~\cite{mr01}).
Thus, besides the obvious IMF effect, the  [$\alpha$/Fe] ratio brings the signature of the timescale 
over which SF has occurred (e.g., Greggio \& Renzini~\cite{grre83}; 
Tornamb\'e \& Matteucci~\cite{tm86}). 
Sub-solar  [$\alpha$/Fe] ratios are explained by chemical-evolution models 
assuming a low rate of SF and a long time-scale for the collapse (e.g., Matteucci~\cite{matt01}).
Lanfranchi \& Matteucci~(\cite{lanfr03}) have shown that a low SF efficiency and a high wind efficiency 
can explain the abundance patterns in dwarf spheroidal galaxies (dSphs) of the Local Group (LG), 
which exhibit highly sub-solar [$\alpha$/Fe] ratios at $[Fe/H]\gsim -1$  (see review by \cite{Tol09} and reference therein).
Marconi et al.~(\cite{marc94}) showed that sub-solar  [$\alpha$/Fe] ratios can be 
obtained with a large number of SF episodes and assuming differential galactic winds that,
being activated only during the bursts where most of the $\alpha$-elements are formed, are more
efficient in depriving the galaxy of the $\alpha$ than of iron.

The presence of an [$\alpha$/Fe] versus $\sigma$ correlation implies that the galaxy gravitational potential well 
regulates the timescale over which the SF occurs.
In fact, the gas evolution within the dark matter halo is controlled by the interplay of 
several factors, such as gravity, radiative cooling, and heating by 
SN feedback, and possibly by an active galactic nucleus (e.g., Chiosi \& Carraro~\cite{cc02}; Granato et al.~\cite{Gra04}). 
In shallower potential wells, SN feedback is increasingly effective in slowing down the SF and driving gas outflows. 
As a consequence, the SF is faster within the most massive haloes.
We also see that the scatter around the age vs. $\sigma$ and $\alpha$/Fe vs. $\sigma$ relations is larger for 
dwarfs than for giants. This was already noticed by Concannon et al.~(\cite{conc00}) and Smith et al.~(\cite{smith09}) using larger 
data samples ($\sim$ 100 and 90 early-type galaxies, respectively). In particular, Concannon et al.~(\cite{conc00}) 
claimed that the increase in the H$\beta$ scatter at the low mass end 
indicates that the smaller galaxies have experienced a more varied star formation history (SFH) and that these 
galaxies have a larger range in age than their brighter counterparts.
This result is consistent with the variety of SFHs  
derived for dSph galaxies in the Local Group from color-magnitude diagram (CMD) fitting: 
ancient ($>$8 Gyr) SF is always detected; however, while some galaxies 
show no evidence of stars younger than 8 Gyr, 
others exhibit prolonged star formations, 
up to few hundred Myrs ago (e.g., Dolphin~\cite{dolph02}, Battaglia et al.~\cite{batt06}, Monelli et al.~\cite{monelli10a,monelli10b}).

We investigated if, besides the galaxy velocity dispersion, the average ages, metallicities, and  [$\alpha$/Fe] ratios  correlate 
with other internal galaxy properties, in the specific, morphology 
(i.e., bulge-to-total light fraction B/T and Sersic index $n$).
We found that the metallicity and the [$\alpha$/Fe] ratio correlate positively with $n$, 
while the correlation of age with $n$ is weak. 
The presence of correlations between metallicity and $n$, and between  [$\alpha$/Fe] and $n$, is expected 
 from the fact that the metallicity and the [$\alpha$/Fe] ratio correlate with $\sigma$, 
 and $\sigma$ correlates with $n$ (see top panel of Figure~\ref{n_afe_bright} and also Trujillo et al.~\cite{Tru04}): 
more massive galaxies tend to be more centrally concentrated, showing higher $n$.
This is observed also if the stellar population parameters are analyzed as function of the bulge-to-total light fraction B/T: the [$\alpha$/Fe] appears to be strongly correlated with B/T, in the sense that the more dominant the bulge, the faster the star formation timescale. 
The presence of a morphology-[$\alpha$/Fe] relation seems in contradiction to the possible evolution along 
the Hubble sequence from low B/T (low $n$) to high B/T (high $n$) galaxies, since it is not possible to produce 
super-solar [$\alpha$/Fe] stars from a gas that is already enriched in iron.
In other words, systems similar to present-day dwarf galaxies couldn't 
be the building blocks of today's giant ellipticals.
This is reminiscent of the problem of the assembly of the MW halo.
The hierarchical formation scenario for the Galactic halo requires the 
accretion and disruption of dwarf galaxies 
(Searle \& Zinn~\cite{sz78}), yet abundance measurements show that low-metallicity halo stars 
are enriched in $\alpha$-elements compared to low metallicity stars 
in dSph galaxies (see review by \cite{Tol09} and reference therein). 
A way to reconcile the observational evidence with a 
hierarchical formation for the halo is to assume that 
it formed at early times from a few relatively massive satellites with 
short-lived, rapid SFHs, and hosting stars enhanced in [$\alpha$/Fe]; 
surviving dSphs, characterized by a slower, less efficient SF, 
are not representative of a typical halo star (Robertson et al.~\cite{robe05}).

Form the study of 48 ETGs from the SAURON sample, Kuntschner et al.~(\cite{Kun10})  suggested that the [$\alpha$/Fe]-$\sigma$ relation is largely driven by an increasing relative contribution with decreasing galaxy luminosity from mildly [$\alpha$/Fe] depressed disk-like components, hosting secondary star formation.
This seems to be in agreement with the fact that we find a strong correlation between [$\alpha$/Fe] and $n$.
However, we do not see a strong correlation between age and $n$, which instead should be observed if 
the contribution from a star forming disk increases with decreasing $n$.
An explanation for this could be that the age is less robustly derived than the [$\alpha$/Fe] ratio, 
because of the large uncertainties in the emission correction to the H$\beta$ index.
Indeed, several studies find a large spread around the age versus $\sigma$ relation, despite the strong 
correlation between [$\alpha$/Fe] and $\sigma$ (e.g., Kuntschner et al~\cite{kunt01}; Thomas et al.~\cite{Tho05}; A07), suggesting that the 
[$\alpha$/Fe] maybe a better measure of the star formation timescales.

Another possibility is that the [$\alpha$/Fe] vs $n$ relation is driven by a progressive variation of the IMF 
with increasing galaxy luminosity/concentration, in the sense that less concentrated galaxies are less  
efficient in producing high mass stars. 
Indeed, some studies (e.g., Hoversten \& Glazebrook~(\cite{hg08}), Meurer et al.~(\cite{meu09})) have claimed the presence of 
systematic variations of the IMF with galaxy properties such as mass, luminosity, and surface brightness.
These studies use large galaxy data samples and are based on the comparison of the H$\alpha$ emission 
with optical color diagrams or with the FUV emission. 
They suggest that low luminosity/ low surface brightness galaxies have less massive stars    
than higher luminosity / higher surface brightness galaxies. 
The idea is that regions of high pressure favor the formation of bound clusters resulting in an increased fraction of 
high mass stars (see Moeckel \& Clarke~(\cite{mc10}) and discussion in Meurer et al.~(\cite{meu09})).

At last, we studied the effect of environment on the evolution of dwarf galaxies.
 First, we investigated,  within the group environment, the effect of local galaxy density 
on the evolution of the stellar populations by 
analyzing  the average ages, metallicities and [$\alpha$/Fe] ratios as a function of 
the local surface galaxy density $\Sigma_3$, based on the distance to the third nearest neighbour. 
We do not find significant correlations with  $\Sigma_3$. 
However, $\Sigma_3$ is a projected surface density, and can be significantly different from the intrinsic 3D galaxy density. 
Projection effects can considerably weaken the correlations with the stellar population parameters.

Our dwarfs, which reside in low density environments (LDE), were also compared with Coma red passive dwarfs already studied in the literature (Smith et al.~\cite{smith09}). 
According to a Mann-Withney test, the age, metallicity, and [$\alpha$/Fe] distributions for the 
Coma core and LDE dwarfs are not significantly different, while a  Kolmogorov-Smirnov test indicates 
that there is a significant difference between the [$\alpha$/Fe] distributions, in the sense that 
LDE dwarfs exhibit lower ratios than Coma dwarfs.
Concluding, we find possible evidence that LDE dwarfs experienced a more prolonged star formation than their 
cluster counterparts, but these results need to be confirmed with larger data samples.

Previous works, based on larger data samples, pointed out a clear effect of local density on the evolution of dwarf galaxies 
in clusters, revealing a trend toward younger mean ages and more extended star formation histories with increasing 
cluster-centric radius (Coma: Poggianti et al.~\cite{pog01a}; Smith et al.~\cite{smith08,smith09};  Abell~496: Chilingarian et al.~\cite{chil08}; Virgo: Michielsen et al.~\cite{mich08}; Toloba et al.~\cite{tolo09}), 
while Paudel, et al.~(\cite{pau10}) found no significant difference in Virgo.
These results are particularly interesting at the light of the fact that no significant difference in [$\alpha$/Fe] is observed 
for giant ETGs in the cluster and in the field (see e.g., Clemens et al.~\cite{Cle09}).
``Suffocation'', ram-pressure stripping, and galaxy harassment may be effective in depriving 
dwarf galaxies from their gas reservoir, thus halting star formation.
Dwarf galaxies residing in low-density environments are likely to retain their gas content and sustain  
star formation for a longer period of time. 

The global scenario that emerges from our study and previous analyses is that 
dwarfs and giants had a different evolution. Giant ETGs seem uniquely affected by their 
own nature (mass) and exhibit old ages and short evolutionary timescales ($\lesssim$ 1 Gyr);
dwarfs present a variety of star formation histories, appear on average younger 
than giants, and seem to be more affected by the environment (nurture).

We can sketch a picture in which today massive ETGs, which completed their star formation process at early times, before the emergence of structures, do not exhibit strong environmental dependence; 
on the other hand, low mass galaxies, which evolved much slower and could eventually feel the effect of environment, 
bring today the sign of {\it nurture}. This scenario is in agreement with findings based on large redshift surveys, which revealed an increasing environmental effect on galaxy evolution moving from higher to lower redshift, and from higher to lower galaxy stellar masses (Iovino et al.~\cite{Iov10}; Peng et al.~\cite{Peng10}, Gr\" utzbauch et al.~\cite{Gru10}).

\begin{acknowledgements}
 F. A. thanks M. Tosi, D. Romano and M. Monelli for useful discussions. 
 We thank the anonymous referee for his/her useful comments which helped to improve the paper.

\end{acknowledgements}

\begin{appendix}

\section{}\label{appendixa}

We give in Table~\ref{taba1} the ages, metallicities, and [$\alpha$/Fe] ratios 
derived for Coma dwarfs reprocessing the H$\beta$, Fe5270, Fe5335, and  Mgb 
indices provided by Smith et al.~(\cite{smith09}) with our own procedure (see Section~3.2).

\begin{table}
\begin{scriptsize}
\caption{Stellar population parameters for Coma dwarfs}\label{taba1}
\begin{tabular}{lccc}
\hline \hline
GMP &  Log Age [yr]  & Log Z/Z$_{\odot}$  &  [$\alpha$/Fe]  \\
\hline
\hline
         4042     &   10.13$\pm$0.17   &   -0.59$\pm$0.18   &  0.22$\pm$0.18     \\
         4035     &     9.86$\pm$0.16   &   -0.33$\pm$0.15   &  0.09$\pm$0.13     \\  
         4029     &   10.08$\pm$0.25   &   -0.64$\pm$0.33   & -0.01$\pm$0.23    \\ 
         4003     &     9.35$\pm$0.10   &    0.07$\pm$0.17   &  0.18$\pm$0.13    \\   
         3973     &     9.49$\pm$0.11   &   -0.05$\pm$0.14   & -0.16$\pm$0.09     \\  
         3969     &     9.84$\pm$0.16   &   -0.44$\pm$0.13   & -0.11$\pm$0.13     \\  
         3855     &     9.84$\pm$0.19   &   -0.23$\pm$0.18   &  0.29$\pm$0.14     \\  
         3780     &     9.85$\pm$0.14   &   -0.19$\pm$0.12   &  0.00$\pm$0.09     \\  
         3719     &   10.03$\pm$0.13   &   -0.22$\pm$0.11   &  0.27$\pm$0.08    \\ 
         3616     &   10.08$\pm$0.22   &   -0.03$\pm$0.21   &  0.02$\pm$0.10    \\ 
         3565     &     9.98$\pm$0.17   &   -0.38$\pm$0.16   & -0.03$\pm$0.13    \\   
         3489     &     9.91$\pm$0.19   &   -0.12$\pm$0.20   &  0.17$\pm$0.12    \\   
         3473     &   10.11$\pm$0.18   &   -0.61$\pm$0.17   &  0.06$\pm$0.15   \\  
         3463     &     9.55$\pm$0.11   &   -0.02$\pm$0.14   &  0.01$\pm$0.08   \\    
         3438     &   10.11$\pm$0.21   &   -0.63$\pm$0.25   & -0.33$\pm$0.17   \\  
         3439     &    9.18$\pm$0.04    &    0.15$\pm$0.06   & -0.01$\pm$0.07    \\   
         3406     &    9.71$\pm$0.16    &   -0.17$\pm$0.16   & -0.06$\pm$0.12    \\   
         3387     &  10.07$\pm$0.17    &   -0.77$\pm$0.21   &  0.10$\pm$0.17    \\ 
         3383     &    9.92$\pm$0.13    &   -0.36$\pm$0.10   &  0.06$\pm$0.09    \\   
         3376     &    9.87$\pm$0.18    &   -0.44$\pm$0.18   &  0.02$\pm$0.16    \\   
         3312     &  10.13$\pm$0.13    &   -0.48$\pm$0.11   &  0.23$\pm$0.11   \\  
         3311     &    9.44$\pm$0.14    &    0.00$\pm$0.21   & -0.08$\pm$0.14   \\    
         3298     &    9.65$\pm$0.12    &   -0.11$\pm$0.11   &  0.14$\pm$0.08   \\    
         3292     &    9.68$\pm$0.11    &   -0.11$\pm$0.09   &  0.03$\pm$0.07   \\    
         3269     &    9.93$\pm$0.23    &   -0.08$\pm$0.30   &  0.19$\pm$0.14   \\    
         3262     &    9.74$\pm$0.14    &    0.25$\pm$0.12   &  0.08$\pm$0.06   \\    
         3248     &    9.86$\pm$0.16    &   -0.46$\pm$0.14   &  0.01$\pm$0.14   \\    
         3238     &    9.56$\pm$0.18    &    0.20$\pm$0.26   &  0.06$\pm$0.09   \\    
         3196     &    9.87$\pm$0.14    &   -0.23$\pm$0.11   & -0.01$\pm$0.09   \\    
         3166     &  10.10$\pm$0.13    &   -0.65$\pm$0.15   &  0.33$\pm$0.17   \\  
         3131     &  10.14$\pm$0.20    &   -0.54$\pm$0.21   & -0.03$\pm$0.16   \\  
         3121     &    9.81$\pm$0.14    &   -0.03$\pm$0.11   & -0.02$\pm$0.08   \\    
         3098     &    9.77$\pm$0.20    &   -0.29$\pm$0.20   & -0.01$\pm$0.16   \\    
         3080     &    9.62$\pm$0.21    &   -0.15$\pm$0.28   &  0.07$\pm$0.17   \\    
         3058     &    9.83$\pm$0.16    &   -0.50$\pm$0.13   &  0.14$\pm$0.15   \\    
         2942     &  10.12$\pm$0.27    &   -0.27$\pm$0.31   &  0.39$\pm$0.17   \\  
         2931     &    9.76$\pm$0.12    &   -0.37$\pm$0.10   &  0.06$\pm$0.11    \\   
         2879     &  10.15$\pm$0.13    &   -0.49$\pm$0.11   &  0.17$\pm$0.11   \\  
         2852     &    9.68$\pm$0.14    &    0.08$\pm$0.12   & -0.09$\pm$0.07    \\   
         2801     &    9.68$\pm$0.20    &   -0.54$\pm$0.24   & -0.06$\pm$0.24    \\   
         2800     &   9.35$\pm$0.14     &    0.03$\pm$0.20   & -0.12$\pm$0.17    \\   
         2799     & 10.12$\pm$0.16     &   -0.54$\pm$0.14   &  0.07$\pm$0.13   \\  
         2784     &   9.86$\pm$0.18     &   -0.26$\pm$0.16   &  0.12$\pm$0.13   \\    
         2764     &   9.58$\pm$0.13     &   -0.20$\pm$0.16   &  0.08$\pm$0.11   \\    
         2753     &   9.60$\pm$0.16     &   -0.26$\pm$0.22   &  0.06$\pm$0.16   \\    
         2728     &   9.92$\pm$0.13     &   -0.31$\pm$0.11   & -0.03$\pm$0.09   \\    
         2692     &   9.62$\pm$0.13     &   -0.23$\pm$0.14   & -0.09$\pm$0.11   \\    
         2676     & 10.08$\pm$0.21     &   -0.74$\pm$0.27   &  0.44$\pm$0.22   \\  
         2626     &   9.71$\pm$0.15     &   -0.19$\pm$0.13   &  0.13$\pm$0.11   \\    
         2591     &   9.97$\pm$0.17     &   -0.52$\pm$0.17   &  0.30$\pm$0.19   \\    
         2585     &   9.60$\pm$0.13     &   -0.34$\pm$0.14   &  0.04$\pm$0.14   \\    
         2529     & 10.05$\pm$0.11     &   -0.40$\pm$0.09   &  0.20$\pm$0.09   \\  
         2478     &   9.81$\pm$0.13     &   -0.30$\pm$0.11   &  0.08$\pm$0.10   \\    
         2420     &   9.37$\pm$0.11     &   -0.15$\pm$0.18   & -0.18$\pm$0.13   \\    
         2411     &  10.11$\pm$0.20    &   -0.59$\pm$0.21   &  0.06$\pm$0.17   \\  
         2376     &   9.44$\pm$0.15     &    0.18$\pm$0.19   &  -0.08$\pm$0.11   \\     
         \hline \\
\end{tabular}
\end{scriptsize}
\end{table}

\end{appendix}


\begin{thebibliography}{}

\bibitem[2007]{Ann07} Annibali, F., Bressan, A., Rampazzo, R., Zeilinger, W. W., Danese, L., 2007, A\&A, 463, 455	

\bibitem[2010]{Ann10}	Annibali, F.; Bressan, A.; Rampazzo, R.; Zeilinger, W. W.; Vega, O.; Panuzzo, P., 2010, eprint arXiv:1004.1647

\bibitem[2009]{bam09} Bamford, S.~P., et al.\ 2009, \mnras, 393, 1324 

\bibitem[2006]{batt06} Battaglia, G., et al.\ 2006, \aap, 459, 423 

\bibitem[2001]{Bek01} Bekki, K., Couch, W. J., Shioya, Y., 2001, PASJ, 53, 395

\bibitem[2002]{Bek02} Bekki, K., Couch, W. J., Shioya, Y., 2002, ApJ, 577, 651
    
\bibitem[1994]{Ber94} Bertelli, G., Bressan, A., Chiosi, C., Fagotto, F., Nasi, E., 1994, A\&AS, 106, 275

\bibitem[1990]{bts90} Binggeli, B., Tarenghi, M., \& Sandage, A.\ 1990, \aap, 228, 42 

\bibitem[1994]{Bre94} Bressan, A., Chiosi, C., Fagotto, F., 1994, ApJS, 94, 63

\bibitem[1996]{bress96} Bressan, A., Chiosi, C., \& Tantalo, R.\ 1996, \aap, 311, 425 

\bibitem[2009]{caffau09} Caffau, E., Maiorca, E., Bonifacio, P., Faraggiana, R., Steffen, M., Ludwig, H.-G., Kamp, I., \& Busso, M.\ 2009, \aap, 498, 877 

\bibitem[2004]{cap04} Cappellari, M., \& Emsellem, E.\ 2004, \pasp, 116, 138 


\bibitem[2008]{chil08} Chilingarian, I.~V., Cayatte, V., Durret, F., Adami, C., Balkowski, C., Chemin, L., Lagan{\'a}, T.~F., \& Prugniel, P.\ 2008, \aap, 486, 85 
   
\bibitem[2001]{Chi01} Chiosi, C., 2001, MmSAI, 72, 733

\bibitem[2002]{cc02} Chiosi, C., \& Carraro, G.\ 2002, \mnras, 335, 335 

\bibitem[2006]{cim06} Cimatti, A., Daddi, E., \& Renzini, A.\ 2006, \aap, 453, L29 

\bibitem[2006]{Cle06} Clemens, M. S., Bressan, A., Nikolic, B., Alexander, P., Annibali, F., Rampazzo, R., 2006, MNRAS, 370, 702

\bibitem[2009]{Cle09} Clemens, M. S., Bressan, A., Nikolic, B.; Rampazzo, R. , 2009, MNRAS, 392, 35

\bibitem[2000]{conc00} Concannon, K.~D., Rose, J.~A., \& Caldwell, N.\ 2000, \apjl, 536, L19 

\bibitem[2006]{DeL06} De Lucia, G., Springel, V., White, S. D. M., Croton, D., Kauffmann, G.,2006, MNRAS, 366, 499
	
\bibitem[2005]{denic05} Denicol{\'o}, G., Terlevich, R., Terlevich, E., Forbes, D.~A., Terlevich, A.\ 2005, \mnras, 358, 813 

\bibitem[2006]{disere06} di Serego Alighieri, S., Lanzoni, B., \& J{\o}rgensen, I.\ 2006, \apjl, 647, L99 

\bibitem[2002]{dolph02} Dolphin, A.~E.\ 2002, \mnras, 332, 91 

\bibitem[1997]{Dre97} Dressler, A., Oemler, A., Jr., Couch, W. J., Smail, I., Ellis, R. S., Barger, A., Butcher, H., Poggianti, B. M., Sharples, R. M., 1997, ApJ, 490, 577

\bibitem[2006]{font06} Fontana, A., et al.\ 2006, \aap, 459, 745 

\bibitem[2007]{ful07} Fulbright, J.~P., McWilliam, A., \& Rich, R.~M.\ 2007, \apj, 661, 1152 

\bibitem[2005]{gall05} Gallazzi, A., Charlot, S., Brinchmann, J., White, S.~D.~M., Tremonti, C.~A.\ 2005, \mnras, 362, 41 

\bibitem[2003]{geha03} Geha, M., Guhathakurta, P., \& van der Marel, R.~P.\ 2003, \aj, 126, 1794 

\bibitem[2008]{gob08} Gobat, R., Rosati, P., Strazzullo, V., Rettura, A., Demarco, R., \& Nonino, M.\ 2008, \aap, 488, 853 

\bibitem[2003]{Gom03} G\'omez, P. L.,, Nichol, R. C., Miller, C. J. et al., 2003, ApJ, 584, 210

\bibitem[1997]{gorg97} Gorgas, J., Pedraz, S., Guzman, R., Cardiel, N., \& Gonzalez, J.~J.\ 1997, \apjl, 481, L19 

\bibitem[1996]{goem96} Goudfrooij, P., \& Emsellem, E.\ 1996, \aap, 306, L45 

\bibitem[2004]{Gra04} Granato, G. L.; De Zotti, G.; Silva, L.; Bressan, A.; Danese, L., 2004, ApJ, 600, 580

\bibitem[1983]{grre83} Greggio, L., \& Renzini, A.\ 1983, Memorie della Societa Astronomica Italiana, 54, 311 

\bibitem[2007]{Gru07} Gr\" utzbauch, R., Trinchieri, G., Rampazzo, R., Held, E.V., Rizzi, L., Sulentic, J.W., Zeilinger, W. W., 2007, AJ, 133, 220

\bibitem[2009]{Gru09} Gr\" utzbauch, R., Zeilinger, W. W., Rampazzo, R., Held, E.V., Sulentic, J.W., Trinchieri, G., 2009, A\&A, 502, 473

\bibitem[2010]{Gru10} Gr\"utzbauch R., Conselice C. J., Varela J., Bundy K., Cooper M. C., Skibba R., Willmer C. N. A., 2010, MNRAS in press, arXiv:1009.3189


\bibitem[2007]{Hai07} Haines, C. P., Gargiulo, A., La Barbera, F., Mercurio, A., Merluzzi, P., Busarello, G., 2007, MNRAS, 381, 7

\bibitem[2007]{haus07} H{\"a}ussler, B., et al.\ 2007, \apjs, 172, 615 

\bibitem[2008]{hg08} Hoversten, E.~A., \& Glazebrook, K.\ 2008, \apj, 675, 163 

\bibitem[1994]{held94} Held, E.~V., \& Mould, J.~R.\ 1994, \aj, 107, 1307 


\bibitem[2010]{Iov10} Iovino A., Cucciati O., Scodeggio M. et al., 2010, A\&A, 509, 40

\bibitem[1995]{Kau95} Kauffmann, G., 1995, MNRAS, 274, 153

\bibitem[1996]{Kau96} Kauffmann, G., 1996, MNRAS, 281, 487

\bibitem[2004]{Kau04} Kauffmann G., White S. D. M., Heckman T. M., M\'enard B., Brinchmann J., Charlot S., Tremonti C., Brinkmann J., 2004, MNRAS, 353, 713

\bibitem[2009]{kol09} Koleva, M., de Rijcke, S., Prugniel, P., Zeilinger, W.~W., \& Michielsen, D.\ 2009, \mnras, 396, 2133 

\bibitem[2005]{Kor05} Korn, A. J., Maraston, C., Thomas, D., 2005, A\&A, 438, 685

\bibitem[2001]{kunt01} Kuntschner, H., Lucey, J.~R., Smith, R.~J., Hudson, M.~J., \& Davies, R.~L.\ 2001, \mnras, 323, 615 

\bibitem[2002]{kunt02} Kuntschner, H., Smith, R.~J., Colless, M., Davies, R.~L., Kaldare, R., \& Vazdekis, A.\ 2002, \mnras, 337, 172 

\bibitem[2010]{Kun10} Kuntschner, H., Emsellem, E., Bacon, R., et al., 2010, eprint arXiv:1006.1574

\bibitem[2010]{laba10} La Barbera, F., Lopes, P.~A.~A., de Carvalho, R.~R., de la Rosa, I.~G., Berlind, A.~A.\ 2010, eprint arXiv:1003.1119 

\bibitem[2003]{lanfr03} Lanfranchi, G.~A., \& Matteucci, F.\ 2003, \mnras, 345, 71 

\bibitem[2003]{LeFevre03} Le F\`evre,O., and the VVDS Team 2003, The Messenger, 111, 18

\bibitem[2002]{Lew02} Lewis, I., Balogh, M., De Propris, R. et al., 2002, MNRAS, 334, 673

\bibitem[2000]{long2000} Longhetti, M., Bressan, A., Chiosi, C., Rampazzo, R.\ 2000, \aap, 353, 917 

\bibitem[1994]{marc94} Marconi, G., Matteucci, F., Tosi, M.\ 1994, \mnras, 270, 35 

\bibitem[2003]{Mar03} Marcolini, A., Brighenti, F., D'Ercole, A., 2003, MNRAS, 345, 1329

\bibitem[2010]{marino10} Marino et al. 2010,  \mnras, accepted

\bibitem[2009]{matk09} Matkovi{\'c}, A., Guzm{\'a}n, R., S{\'a}nchez-Bl{\'a}zquez, P., Gorgas, J., Cardiel, N., Gruel, N.\ 2009, \apj, 691, 1862 

\bibitem[2001]{mr01} Matteucci, F., \& Recchi, S.\ 2001, \apj, 558, 351 


\bibitem[2001]{matt01} Matteucci, Book Review: The chemical evolution of the galaxy / Kluwer, 2001

\bibitem[Mayer et al. (2006)]{May06} Mayer, L., Mastropietro, C., Wadsley, J., Stadel, J., Moore, B., 2006, MNRAS, 369, 1021


\bibitem[2009]{meu09} Meurer, G.~R., et al.\ 2009, \apj, 695, 765 

\bibitem[2008]{mich08} Michielsen, D., et al.\ 2008, \mnras, 385, 1374 

\bibitem[2010]{mc10} Moeckel, N., \& Clarke, C.~J.\ 2010, arXiv:1009.0283 

\bibitem[2010b]{monelli10b} Monelli, M., et al.\ 2010, \apj, 722, 1864 

\bibitem[2010a]{monelli10a} Monelli, M., et al.\ 2010, \apj, 720, 1225 


\bibitem[2005]{nel05} Nelan, J.~E., Smith, R.~J., Hudson, M.~J., Wegner, G.~A., Lucey, J.~R., Moore, S.~A.~W., 
Quinney, S.~J., \& Suntzeff, N.~B.\ 2005, \apj, 632, 137 

\bibitem[1974]{Oem74} Oemler A. Jr., 1974, ApJ, 194, 10
     
\bibitem[1995]{Oh95} Oh, K. S., Lin, D. N. C., Aarseth, S. J., 1995, ApJ, 442, 142

\bibitem[2005]{ott05} Ott, J., Walter, F., \& Brinks, E.\ 2005, \mnras, 358, 1453 

\bibitem[2009]{pan09} Pannella, M., et al.\ 2009, \apj, 701, 787 

\bibitem[2010]{pau10} Paudel, S., Lisker, T., Kuntschner, H., Grebel, E.~K., \& Glatt, K.\ 2010, \mnras, 570 

\bibitem[2002]{Peng02} Peng, C. Y., Ho, L. C., Impey, C. D. \& Rix, H., 2002, AJ, 124, 266

\bibitem[2010]{Peng10} Peng, Y., et al.\ 2010, arXiv:1003.4747 

\bibitem[2001]{pog01a} Poggianti, B.~M., et al.\ 2001, \apj, 562, 689 

\bibitem[2010]{pog10} Poggianti, B.~M., De Lucia, G., Varela, J., Aragon-Salamanca, A., Finn, R., Desai, V., von der 
Linden, A., \& White, S.~D.~M.\ 2010, \mnras, 405, 995 


\bibitem[2004]{proct04} Proctor, R.~N., Forbes, D.~A., Hau, G.~K.~T., Beasley, M.~A., De Silva, G.~M., Contreras, R., Terlevich, A.~I.\ 2004, \mnras, 349, 1381 

\bibitem[2001]{pru01} Prugniel, P., \& Soubiran, C.\ 2001, \aap, 369, 1048 

\bibitem[2005]{Ram05} Rampazzo, R., Annibali, F., Bressan, A., Longhetti, M., Padoan, F., Zeilinger, W. W., 2005, A\&A, 433, 497

\bibitem[2001]{recchi01} Recchi, S., Matteucci, F., D'Ercole, A.\ 2001, \mnras, 322, 800 


\bibitem[2007]{reda07} Reda, F.~M., Proctor, R.~N., Forbes, D.~A., Hau, G.~K.~T., Larsen, S.~S.\ 2007, \mnras, 377, 1772 

\bibitem[1995]{ReRa95} Reduzzi, L. \& Rampazzo, R., 1995, ApLC, 30, 1

\bibitem[2010]{rett10} Rettura, A., et al.\ 2010, \apj, 709, 512 

\bibitem[2005]{robe05} Robertson, B., Bullock, J.~S., Font, A.~S., Johnston, K.~V., \& Hernquist, L.\ 2005, \apj, 632, 872 


\bibitem[1988]{Rub88} Rubin V. C., Whitmore B. C., Ford W. K. Jr., 1988, ApJ, 333, 522

\bibitem[2006]{sb06} S{\'a}nchez-Bl{\'a}zquez, P., Gorgas, J., Cardiel, N., \& Gonz{\'a}lez, J.~J.\ 2006, \aap, 457, 809 

\bibitem[1976]{sche76} Schechter, P.\ 1976, \apj, 203, 297 

\bibitem[1978]{sz78} Searle, L., \& Zinn, R.\ 1978, \apj, 225, 357 

\bibitem[2008]{smith08} Smith, R.~J., et al.\ 2008, \mnras, 386, L96 

\bibitem[2009]{smith09} Smith, R.~J., Lucey, J.~R., Hudson, M.~J., Allanson, S.~P., Bridges, T.~J., Hornschemeier, 
A.~E., Marzke, R.~O., \& Miller, N.~A.\ 2009, \mnras, 392, 1265 

\bibitem[2010]{spo10} Spolaor, M., Kobayashi, 
C., Forbes, D.~A., Couch, W.~J., \& Hau, G.~K.~T.\ 2010, arXiv:1006.1698 

\bibitem[2009]{Tas09} Tasca L. A. M., Kneib J.-P., Iovino A. et al., 2009, A\&A, 503, 379

\bibitem[2003]{Tho03} Thomas, D., Maraston, C., Bender, R., 2003, MNRAS, 339, 89

\bibitem[2005]{Tho05} Thomas, D., Maraston, C., Bender, R., Mendes de Oliveira, C., 2005, ApJ., 621, 673

\bibitem[2010]{Tho10} Thomas, D., Maraston, C., Schawinski, K., Sarzi, M., \& Silk, J.\ 2010, \mnras, 404, 1775 

\bibitem[2009]{tolo09} Toloba, E., et al.\ 2009, \apjl, 707, L17 
 
\bibitem[Tolstoy et al. (2009)]{Tol09} Tolstoy, E., Hill, V., Tosi, M., 2009, ARA\&A, 47, 371

\bibitem[1986]{tm86} Tornamb{\'e}, A., \& Matteucci, F.\ 1986, \mnras, 223, 69 

\bibitem[1998]{Tra98} Trager, S. C., Worthey, G., Faber, S. M., Burstein, D., Gonzalez, J. J., 1998 ApJS, 116, 1
   
\bibitem[2000]{Tra00} Trager, S. C., Faber, S. M., Worthey, G., González, J. J., 2000, AJ, 119, 1645

\bibitem[2001]{TR01} Trinchieri, G. \& Rampazzo, R., 2001, AA, 374, 454

\bibitem[2004]{Tru04} Trujillo, I., Burkert, A., \& Bell, E.~F., 2004, \apjl, 600, L39

\bibitem[1988a]{Tul88} Tully, R. B., 1988, Nearby Galaxy Catalogue, Cambridge University Press

\bibitem[1988b]{tully88b} Tully, R.~B.\ 1988, \aj, 96, 73 

\bibitem[2007]{vando07} van Dokkum, P.~G., \& van der Marel, R.~P.\ 2007, \apj, 655, 30 

\bibitem[2004]{vanz04} van Zee, L., Barton, E.~J., \& Skillman, E.~D.\ 2004, \aj, 128, 2797 

\bibitem[1978]{Whi78} White, S. D. M., Rees, M. J., MNRAS, 183, 341

\bibitem[1994]{worth94} Worthey, G., Faber, S.~M., Gonzalez, J.~J., \& Burstein, D.\ 1994, \apjs, 94, 687 
   
\bibitem[1997]{Wor97} Worthey, G., Ottaviani, D. L., 1997, ApJS, 111, 377


\end{thebibliography}
\end{document}